\documentclass[
universe,
article,
submit,
moreauthors]{Definitions/mdpi} 

\firstpage{1} 
\pubvolume{1}
\issuenum{1}
\articlenumber{0}
\pubyear{2021}
\copyrightyear{2020}
\datereceived{} 
\dateaccepted{} 
\datepublished{} 
\hreflink{https://doi.org/} 

\usepackage{xcolor}
\usepackage{dcolumn}
\usepackage{lipsum}
\usepackage[caption=false]{subfig}

\usepackage[T1]{fontenc}

\preto{\abstractkeywords}{\nolinenumbers}

\Title{High-Order Multipole and Binary Love Number Universal Relations}

\TitleCitation{Title}

\Author{Daniel A. Godzieba$^{1}$ \orcidA{} and David Radice$^{2,1,3}$ \orcidB{}}

\AuthorNames{Daniel A. Godzieba, David Radice}

\AuthorCitation{Godzieba, D. A. and Radice, D.}

\address{%
$^{1}$ \quad Department of Physics, The Pennsylvania
State University, University Park, Pennsylvania 16802\\
$^{2}$ \quad Institute for Gravitation \& the Cosmos, The Pennsylvania State University, University Park, PA 16802 \\
$^{3}$ \quad Department of Astronomy \& Astrophysics, The Pennsylvania State University, University Park, PA 16802}

\corres{Correspondence: dag5611@psu.edu}

\abstract{Using a data set of approximately 2 million phenomenological equations of state consistent with observational constraints, we construct new equation-of-state-insensitive universal relations that exist between the multipolar tidal deformability parameters of neutron stars, $\Lambda_l$, for several high-order multipoles ($l = 5,6,7,8$), {and we consider finite-size effects of these high-order multipoles in waveform modelling.} We {also} confirm the existence of a universal relation between the radius of the $1.4 M_\odot$ NS, $R_{1.4}$ and the reduced tidal parameter of the binary, $\tilde{\Lambda}$, and the chirp mass. We extend this relation to a large number of chirp masses and to the radii of isolated NSs of different mass $M$, $R_M$. We find that there is an optimal value of $M$ for every $\mathcal{M}$ such that the uncertainty in the estimate of $R_M$ is minimized when using the relation. We discuss the utility and implications of these relations for the upcoming LIGO O4 run and third-generation detectors.}


\begin{document}
\section{Introduction}

Due to the constraints imposed by general relativity and causality, there exist quasi-universal relations between various bulk physical properties of neutron stars (NSs) that are mostly insensitive to the actual equation of state (EOS) of nuclear matter \cite{Andersson:1996pn,Andersson:1997rn,Benhar:1998au,Benhar:2004xg,Tsui:2004qd,Lau:2009bu,Bejger:2002ty,Lattimer:2004nj,Urbanec:2013fs,Yagi:2013bca,Yagi:2013awa,Yagi:2013sva,Yagi:2014qua,Godzieba:2020bbz}. Since the nuclear EOS in the high-density regime of NSs is still unknown, these universal relations are a great utility for gravitational wave (GW) astronomy. Universal relations reduce a group of several seemingly independent physical properties to a family characterized by only a few parameters. Ideally, this allows one to break the degeneracies between parameters in the analysis of GW data as well as in waveform modelling.

A robust set of universal relations (called multipole Love relations) holds between the $l$-th order dimensionless gravitoelectric tidal deformability coefficients of NSs \cite{Yagi:2013sva}, $\Lambda_l$, which are defined by \begin{equation}
    \Lambda_l \equiv \frac{2}{(2l-1)!!} \frac{k_l}{C^{2l+1}},
    \label{Lambda definition}
\end{equation} where $C = M/R$ is the compactness of the NS (here we take $G = c = 1$) and $k_l$ is its $l$-th order gravitoelectric tidal Love number \cite{Hinderer:2009ca}. The GW waveform of a binary NS (BNS) merger is, quite understandably, highly sensitive to these tidal parameters. How deformable a NS is in a tidal potential affects how its mass ultimately gets distributed during the inspiral of a merger, which, in turn, shapes the GW waveform, especially during the late stages of the inspiral \cite{Hinderer:2007mb,Hinderer:2009ca,Damour:2009vw,Nagar:2018plt,Zhao:2018nyf}. The tidal parameters enter into the waveform at different post-Newtonian orders; however, they are degenerate in the signal \cite{Yagi:2013sva}. The multipole relations allow this degeneracy to be broken by reducing all of the tidal deformabilities to a family determined by a single parameter. This parameter is always chosen to be the quadrupolar tidal deformability $\Lambda_2$, which is the source of the leading-order finite-size effect in the GW signal and, consequently, is the easiest to measure \cite{Yagi:2013sva}. Thus, higher-order ($l>2$) tidal deformabilities can be expressed through the multipole Love relations as functions of $\Lambda_2$. The authors and others have demonstrated that the improvements to the accuracy of tidal deformability measurements, to parameter estimation, and to GW modelling offered by the multipole Love relations are significant \cite{Godzieba:2020bbz,Yagi:2013sva,Yagi:2016bkt,Hinderer:2009ca,Gamba:2020wgg} and will become particularly important with the increased sensitivity of upcoming third-generation GW detectors like LIGO III, the Einstein Telescope, and Cosmic Explorer \cite{Yagi:2013sva,Hinderer:2009ca,Adhikari:2013kya,ET-GW,Hild:2008ng,Reitze:2019iox}.

Motivated by these potential improvements, we present entirely new fits to several previously un-fitted high-order multipole Love relations, specifically for $l = 5,6,7,$ and $8$. Though the finite-size effects of these orders of tidal parameters are currently smaller than measurement error, they will become more measurable with increased sensitivity; hence, faithful GW waveform modelling will need to incorporate them. Previosu studies, such as \citet{Flanagan:2007ix} and \citet{Damour:2012yf}, have discussed the finite-size effects of the $l \leq 4$ multipoles.

\citet{Zhao:2018nyf, De:2018uhw} have demonstrated the existence of an intriguing EOS-insensitive relation for BNSs between the radius of the $1.4 M_\odot$ NS, $R_{1.4}$ and the reduced tidal deformability (also called the binary tidal deformability), $\tilde{\Lambda}$. The quadrupolar deformabilities of the individual NSs enter into the GW signal of the merger via $\tilde{\Lambda}$, which is defined as \begin{equation}
    \tilde{\Lambda} \equiv \frac{16}{13} \frac{(12q+1)\Lambda_{2,1} + (12+q)\Lambda_{2,2}}{(1+q)^5}, \label{tildeLambda}
\end{equation} where $\Lambda_{2,1}$ and $\Lambda_{2,2}$ are the deformabilities of the primary and the secondary stars respectively. The quadrupolar tidal Love number $k_2$ is known to scale roughly as $C^{-1}$ independently of the EOS \cite{Hinderer:2009ca,Postnikov:2010yn}. According to Eq. (\ref{Lambda definition}), this means $\Lambda_2$ scales approximately as $C^{-6}$. In an apparently analogous fashion, $\tilde{\Lambda}$ seems to go as $(\mathcal{M}/R_{1.4})^{-6}$, where $\mathcal{M}$ is the chirp mass of the BNS given by \begin{equation}
    \mathcal{M} \equiv \frac{(m_1 m_2)^{3/5}}{(m_1 + m_2)^{1/5}}. \label{Mchirp definition}
\end{equation} Combining this observation with the definition of $\tilde{\Lambda}$ in the manner done by \citet{Zhao:2018nyf} yields a mostly EOS-insensitive estimate of $R_{1.4}$ in terms of $\tilde{\Lambda}$ and $\mathcal{M}$ that is also mostly insensitive to the binary mass ratio $q$: \begin{equation}
    R_{1.4} \simeq (11.5 \pm 0.3 \ {\rm km}) \frac{\mathcal{M}}{M_\odot} \left( \frac{\tilde{\Lambda}}{800} \right)^{1/6}. \label{R-LT relation}
\end{equation}

The immediate utility of this relation is the ability to produce an EOS-agnostic estimate of $R_{1.4}$ from just tidal parameter measurements. This is an alternative to the more involved method of using the universal relation for binaries between the symmetric and antisymmetric combinations of $\Lambda_{2,1}$ and $\Lambda_{2,2}$ \cite{Yagi:2015pkc,Godzieba:2020bbz} combined with the relation for individual NSs between $\Lambda_2$ and the compactness $C$ \cite{Maselli:2013mva,Godzieba:2020bbz} (a relation which intuitively follows from the definition of $\Lambda_l$ in Eq. (\ref{Lambda definition})). One would first use the symmetric-antisymmetric relation to break the degeneracy between $\Lambda_{2,1}$ and $\Lambda_{2,2}$ and estimate them individually from $\tilde{\Lambda}$, and then use the $\Lambda_2$-$C$ relation and the masses of the binary to extract the radii of both stars. The LIGO/VIRGO analysis of GW170817 is an example of this latter approach \cite{TheLIGOScientific:2017qsa,Abbott:2018exr,De:2018uhw}. One need not appeal to universal relations to estimate stellar radii, however. Instead, one could perform an inference of the EOS directly using a parametric representation of the EOS, as was also done in the LIGO/VIRGO analysis \cite{TheLIGOScientific:2017qsa}, or using a much more sophisticated nonparametric representation, as described in \citet{Essick:2019ldf}.

It is an appealing question, then, whether this relation can be extended using the radius of a NS with a generic mass $M$, $R(M) = R_M$. A $R_M$-$\tilde{\Lambda}$ relation would allow one to use measurements of tidal parameters and $\mathcal{M}$ to place robust constrains on $R_M$ directly without the need for a more complicated procedure. Hence, our motivation in this work is to provide a phenomenological study of the $R_M$-$\tilde{\Lambda}$ relation. We look at the relation for several values of $M$. For a given $M$, we compute fits to the relation for twelve fixed values of $\mathcal{M}$ between $0.9 M_\odot$ and $1.4 M_\odot$. We then generalize the fit for all $\mathcal{M} \in [0.9 M_\odot, 1.4 M_\odot]$ by interpolating the fitting parameters as functions of $\mathcal{M}$. Fitting to the relation from across a vast set of phenomenological EOSs incorporates the effects of higher-order terms that are dropped when one analytically derives the expression in Eq. (\ref{R-LT relation}) as was done in \cite{Zhao:2018nyf}. Eq. (\ref{R-LT relation}) assumes that the $R_{1.4}$-$\tilde{\Lambda}$ relation (and, by extension, the $R_M$-$\tilde{\Lambda}$ relation) is only linearly dependent on $\mathcal{M}$, i.e. that $\mathcal{M}$ simply scales the relation but does not change its dependence on $\tilde{\Lambda}$. A phenomenological study permits us to observe directly the effect changing $\mathcal{M}$ has on the relation.

The outline of this paper is as follows. In Sec. \ref{Methods}, we describe the parameterization scheme and algorithm by which we generate our phenomenological EOS data and the statistics for our analyses. In Sec. \ref{High-Order Multipole Relations}, we present the fitting parameters of the high-order multipole Love relations, followed in Sec. \ref{R-LT Relation} by an phenomenological analysis of and fits to the $R_{1.4}$-$\tilde{\Lambda}$ relation as well as to the general $R_{M}$-$\tilde{\Lambda}$ relation. We also discuss the implications of these new fits to GW waveform analysis for the LIGO O4 run. A concluding summary is given in Sec. \ref{Conclusions}.

\section{\label{Methods}Methods}

We parameterize the space of all possible EOSs consistent with theoretical calculations and astronomical observations using the piecewise polytropic interpolation developed in \citet{Read:2008iy}, with the only modification being that we allow the transition densities $\rho_1$ and $\rho_2$ to vary. We then generate random piecewise EOSs using a Markov chain Monte Carlo (MCMC) algorithm, with the basic summary as follows. For a given candidate EOS, the algorithm first computes a series of solutions to the Tolman-Oppenheimer-Volkoff (TOV) equation using the publically available \texttt{TOVL} code described in \citet{Bernuzzi:2008fu} and \citet{Damour:2009vw}, and then accepts the EOS if and only if it satisfies three weak physical constraints: \begin{enumerate}
    \item causality of the maximum mass NS is preserved (i.e. the maximum sound speed $c_s$ is less than the speed of light $c$ below the maximum stable central density),
    \item the maximum stable mass of a non-rotating NS, $M_{\rm max}$, is greater than $1.97 M_\odot$, and
    \item $\Lambda_2 < 800$ for the $1.4 M_\odot$ NS.
\end{enumerate} The full details of parameterization and the MCMC algorithm can be found in \citet{Godzieba:2020bbz}. With this scheme, we generate a set of 1,966,225 phenomenological EOSs.

To study the multipole Love relations, for each EOS in our data set, we solve the TOV equation for sixteen evenly spaced central densities between $\rho_c = 3.09\times10^{14} \ {\rm g}/{\rm cm}^3$ and the maximum stable central density of that EOS, and then extract $\Lambda_l$ for $l = 2$ through $l = 8$ from each solution.

To study the $R_{M}$-$\tilde{\Lambda}$ relation, we follow a similar procedure. First, we choose a fixed value of $\mathcal{M}$. Next, for each EOS in a random sample of a quarter of all EOS in the data set, we generate twenty random binary NSs (BNSs). We uniformly sample the binary mass ratio $q = m_2 / m_1$ (where $m_2 \leq m_1$) on the interval $1/2 \leq q \leq 1$. This range is not intended to represent the complete range of values that $q$ could take in Nature, but rather simply to capture the general behavior of $q$ based on observational and theoretical considerations. Observations of the most massive known pulsars indicate that $M_{\rm max} \gtrsim 2 M_\odot$ \cite{Demorest:2010bx,Fonseca:2016tux,Arzoumanian:2017puf,Antoniadis:2013pzd,Cromartie:2019kug,Linares:2018ppq}, and the analysis of GW170817 suggests that $M_{\rm max} \lesssim 2.3M_\odot$ \cite{Margalit:2017dij,Shibata:2017xdx,Rezzolla:2017aly,Ruiz:2017due,Shibata:2019ctb}; though, as we await upcoming precision measurements of millisecond pulsar radii by NICER, we cannot as of yet categorically rule out the possibility of extreme EOSs with $M_{\rm max} < 2.5M_\odot$ \cite{Godzieba:2020tjn}. Meanwhile, the least massive known pulsar has a mass of $1.17M_\odot$ \cite{Martinez:2015mya}, and, depending on the true nuclear EOS, the minimum stable gravitational mass, $M_{\rm min}$, could be as low as $1.15M_\odot$ \cite{Suwa:2018uni}. Hence, $q \geq M_{\rm min} / M_{\rm max} \approx 1/2$, and the vast majority of BNSs, being far from either mass extreme, will fall well within this range.

Each $q$ is then converted into the actual binary masses $m_1$ and $m_2$ using the value of $\mathcal{M}$ and Eq. (\ref{Mchirp definition}): \begin{equation}
    m_1 = \mathcal{M} q^{-3/5}(1+q)^{1/5}, \qquad  m_2 = \mathcal{M} q^{2/5}(1+q)^{1/5}. \label{m1 and m2 definitions}
\end{equation} The TOV equation is then solved with the corresponding EOS for NSs with these two masses, and $\Lambda_2$ is extracted from both solutions to compute $\tilde{\Lambda}$. We apply this procedure for twelve different values of $\mathcal{M}$ between $0.9 M_\odot$ and $1.4 M_\odot$. The $\tilde{\Lambda}$ values are then plotted versus $R_M$ for eight different values of $M$. These $R_M$ values are pulled from our EOS data set.

\section{\label{High-Order Multipole Relations}High-Order Multipole Relations}

\begin{figure}
	\centering
	\includegraphics[trim=50 0 50 0, clip, width=0.6\linewidth]{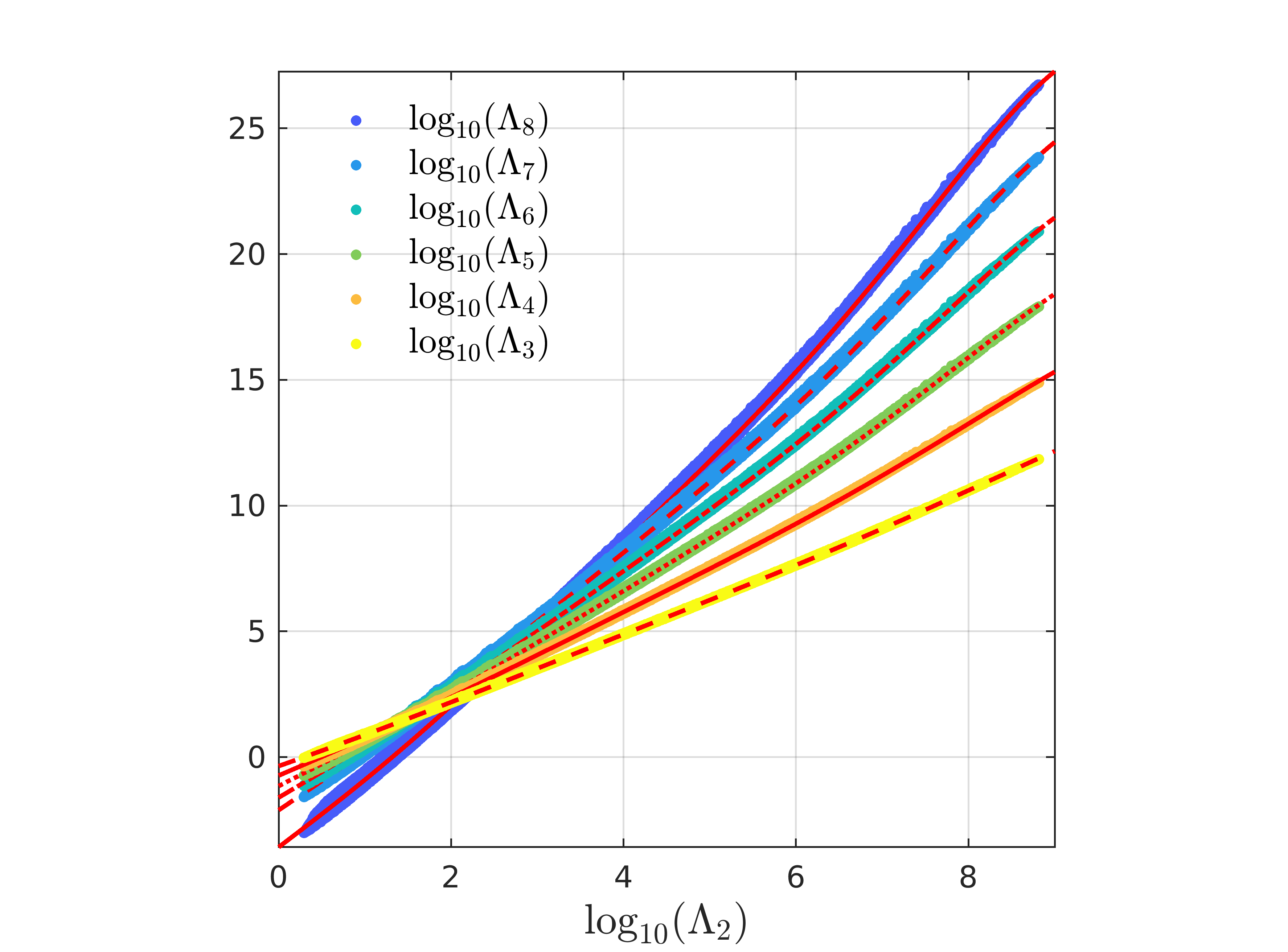}
	\caption{Universal multipole Love relations for $l = 3$ through $l = 8$ from the collection of phenomenological EOSs. We use the fitting function function in Eq. (\ref{multipole fit}), and the fit to each relation is plotted in red.}
	\label{high-order multipoles}
\end{figure}

\begin{figure*}[t]
    \centering
    \resizebox{\linewidth}{!}{%
    \subfloat[]{{\includegraphics[trim=45 0 70 0, clip, width=0.45\linewidth]{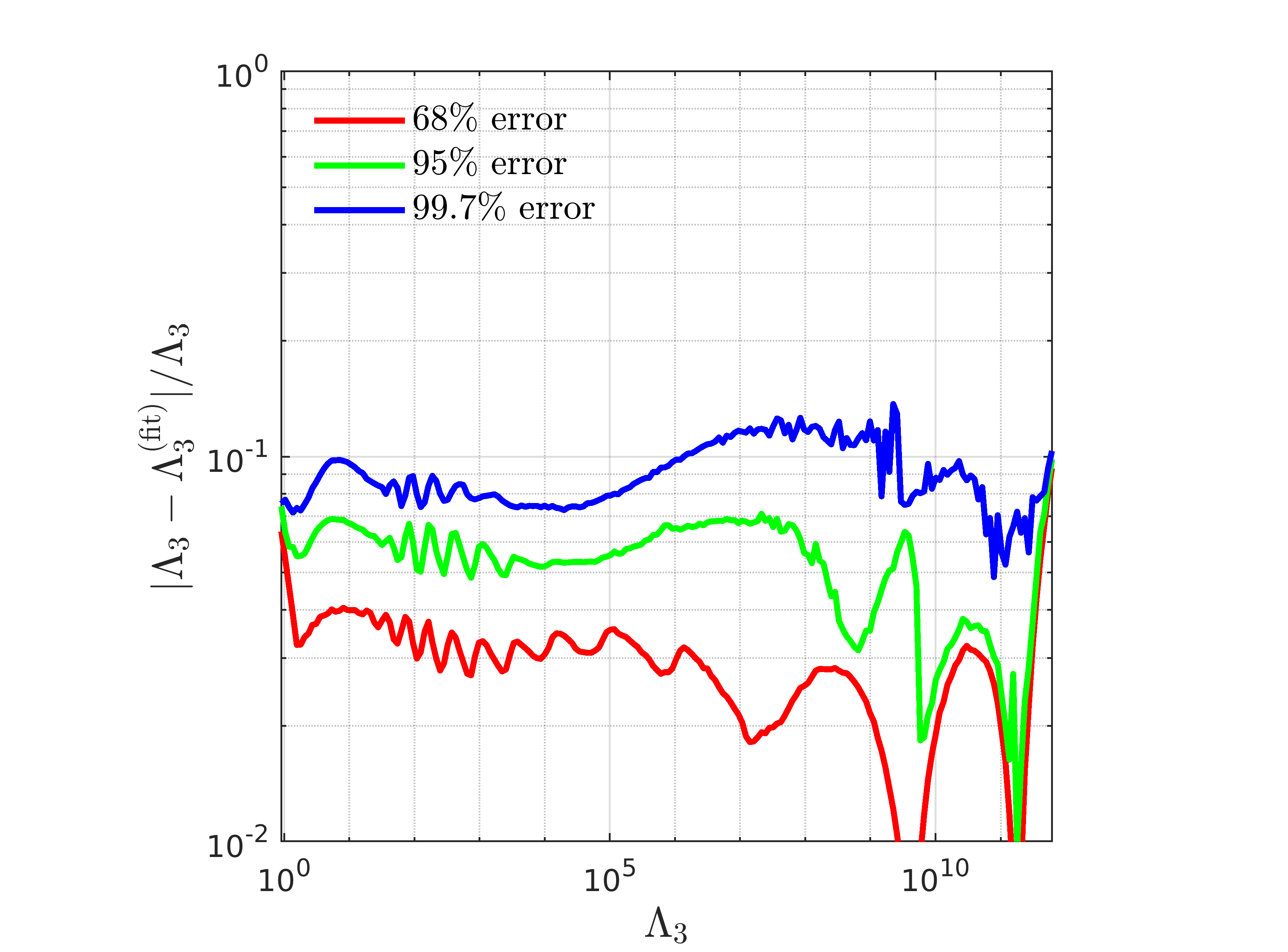}}}%
    \quad
    \subfloat[]{{\includegraphics[trim=45 0 70 0, clip, width=0.45\linewidth]{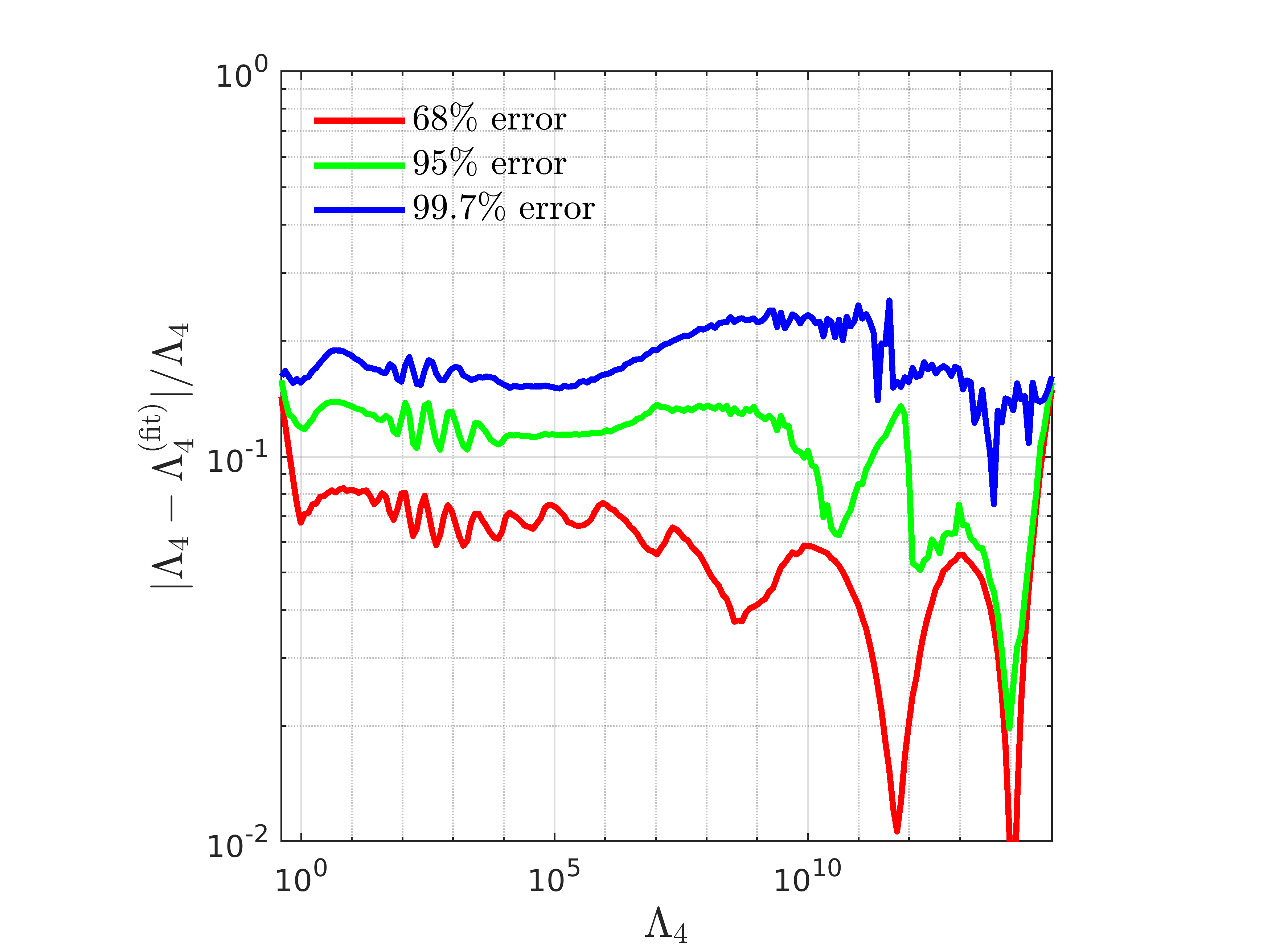}}}%
    \quad
    \subfloat[]{{\includegraphics[trim=45 0 70 0, clip, width=0.45\linewidth]{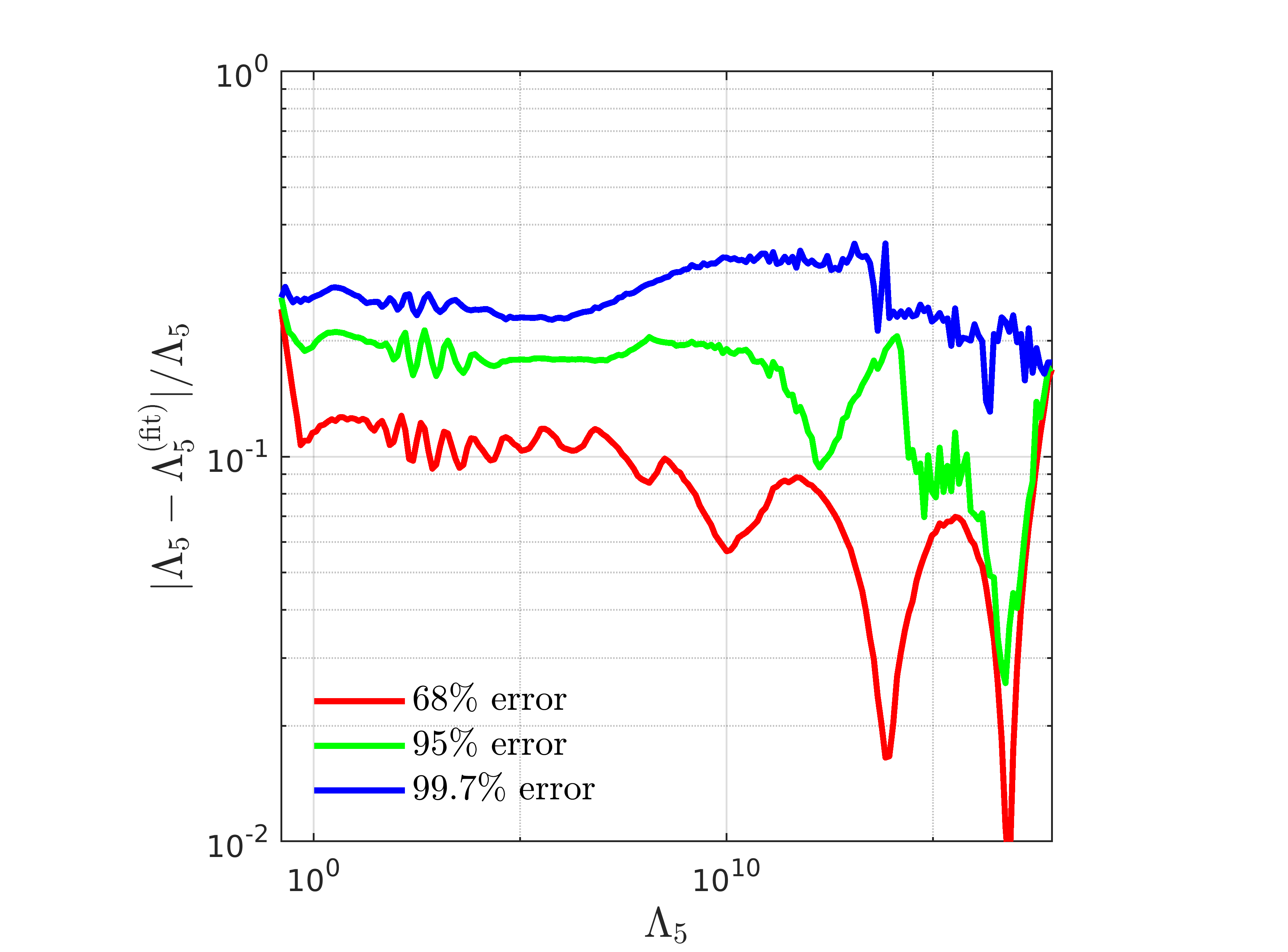}}}}\\
    \resizebox{\linewidth}{!}{%
    \subfloat[]{{\includegraphics[trim=45 0 70 0, clip, width=0.45\linewidth]{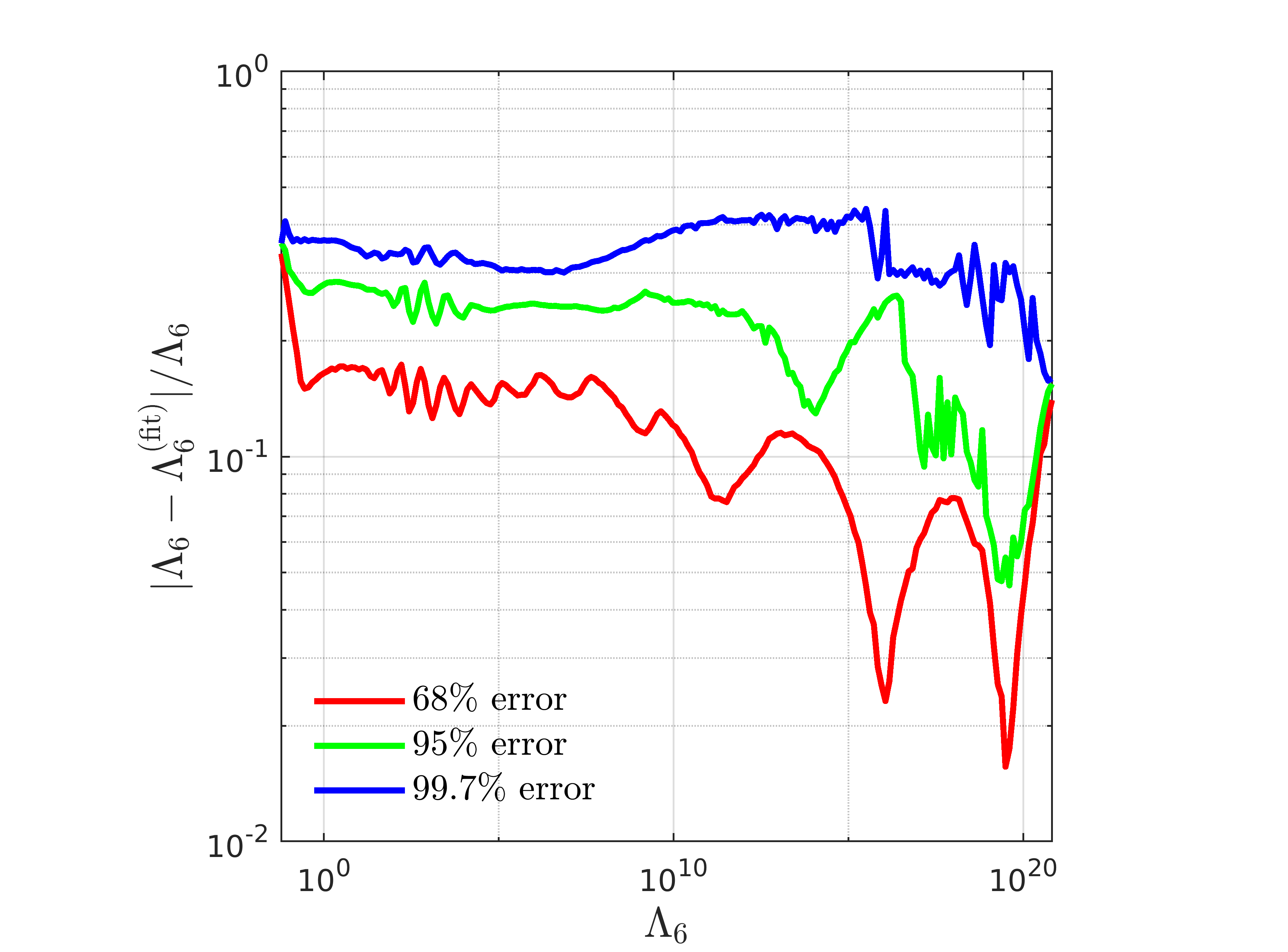}}}%
    \quad
    \subfloat[]{{\includegraphics[trim=45 0 70 0, clip, width=0.45\linewidth]{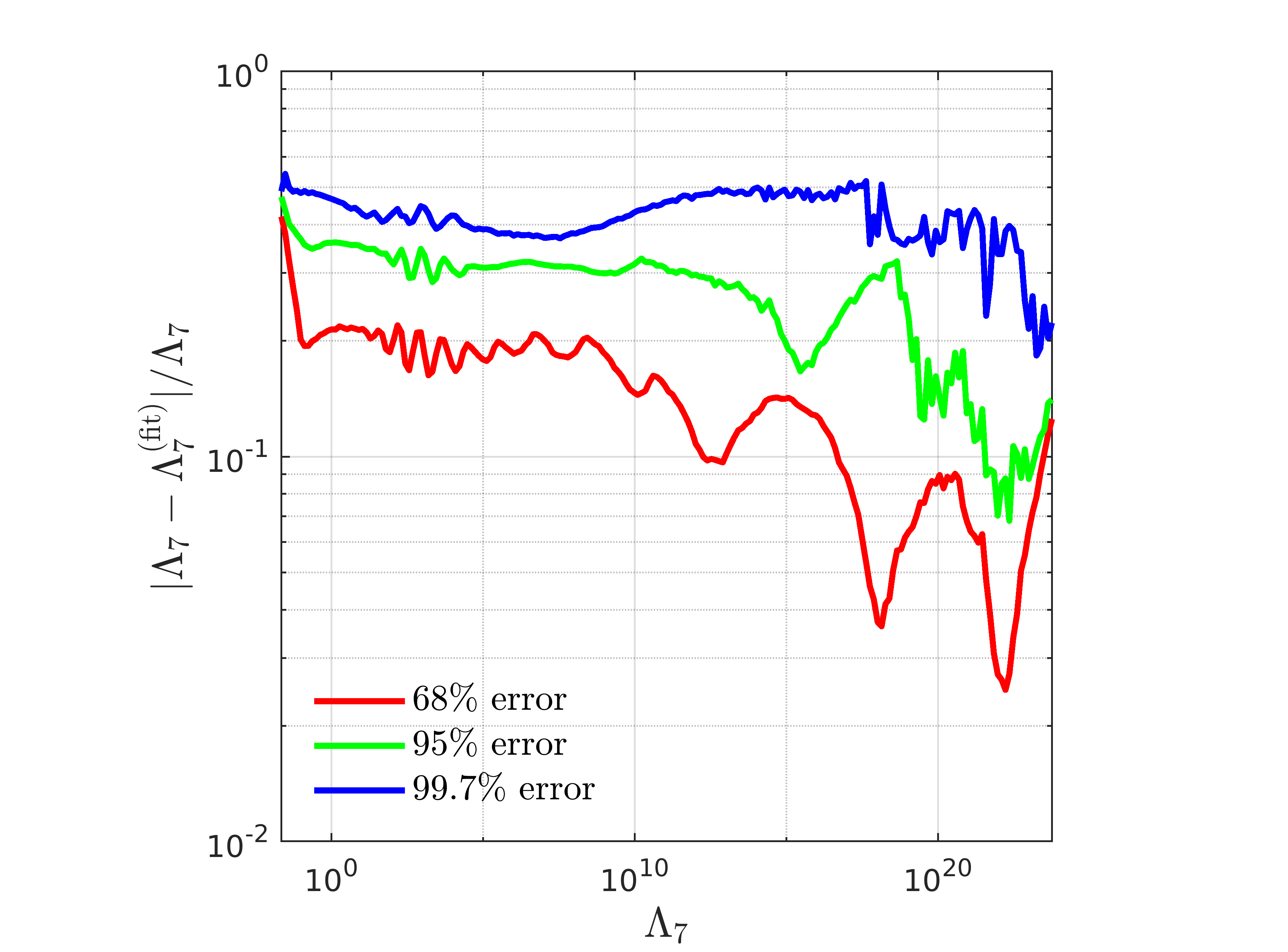}}}%
    \quad
    \subfloat[]{{\includegraphics[trim=45 0 70 0, clip, width=0.45\linewidth]{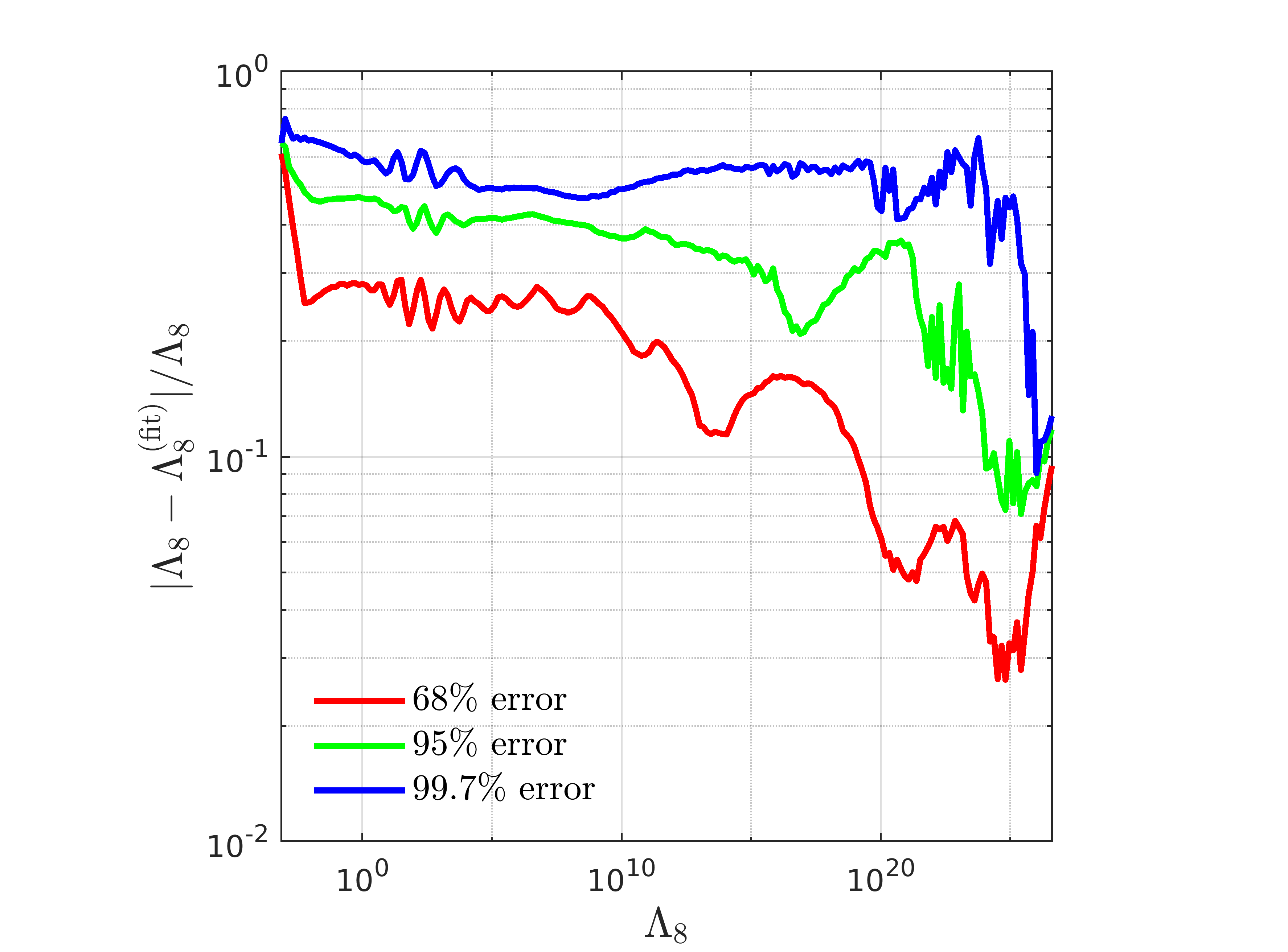}}}}%
    \caption{68\%, 95\%, and 99.7\% relative errors of the fits to {(a) $\Lambda_3$-$\Lambda_2$, (b) $\Lambda_4$-$\Lambda_2$, (c) $\Lambda_5$-$\Lambda_2$, (d) $\Lambda_6$-$\Lambda_2$, (e) $\Lambda_7$-$\Lambda_2$, and (f) $\Lambda_8$-$\Lambda_2$ relations}. The small ripple in the error seen at small values of each $\Lambda_l$ is simply an artifact of how the distribution of NS solutions were generated. The fits are faithful to the shape of the curves of the relations; however, universality weakens and the distributions of points spread out as $l$ increases, resulting in the maximum error of the estimate increasing with $l$.}%
    \label{multipole error}%
\end{figure*}

From our phenomenological EOS data set, we compute 21,994,104 valid individual NS solutions to the TOV equation as our statistics for analyzing the multipole Love relations. $\Lambda_3$, $\Lambda_4$, $\Lambda_5$, $\Lambda_6$, $\Lambda_7$, and $\Lambda_8$ are plotted against $\Lambda_2$ in Fig. \ref{high-order multipoles}, and one can appreciate the universality of each relation across a vast range of scales. (We observe that all the intersections between the six curves lie around $\Lambda_2 \sim 100$, but we are not sure why this is the case.) As in the authors' previous work \cite{Godzieba:2020bbz}, we employ a fitting function of the form \begin{equation}
    \ln{\Lambda_{l}} = \sum_{k=0}^{6} a_{k} \left( \ln{\Lambda_{2}} \right)^{k},
    \label{multipole fit}
\end{equation} which is an extended version of the fitting function originally used by \citet{Yagi:2013awa}. The fitting parameters $\vec{a} = \{a_k\}$ for each relation are given in Table \ref{multipole fitting parameters}. \begin{specialtable}[b]
  \caption{Fitting parameters $\vec{a} = \{a_k\}$ of the multipole Love relations given in Eq. (\ref{multipole fit}).}
  \centering
  \resizebox{\linewidth}{!}{%
  \begin{tabular}{cccccccc}
    \hline \hline
      Relation   & $a_{0}$ & $a_{1}$ & $a_{2}$ & $a_{3}$ & $a_{4}$ & $a_{5}$ & $a_{6}$ \\
    \hline
      $\Lambda_3$-$\Lambda_2$   &  $-0.82195$  &  $1.2110$  &  $1.0494\times10^{-2}$  &  $1.6581\times10^{-3}$  &  $-3.1933\times10^{-4}$  &  $1.8607\times10^{-5}$  &  $-3.5027\times10^{-7}$ \\
      $\Lambda_4$-$\Lambda_2$   &  $-1.6887$  &  $1.4719$  &  $7.1803\times10^{-3}$  &  $5.4042\times10^{-3}$  &  $-8.3262\times10^{-4}$  &  $4.6940\times10^{-5}$  &  $-8.9092\times10^{-7}$ \\
      $\Lambda_5$-$\Lambda_2$   &  $-2.6473$  &  $1.7485$  &  $-5.1199\times10^{-4}$  &  $9.7085\times10^{-3}$  &  $-1.3990\times10^{-3}$  &  $7.8465\times10^{-5}$  &  $-1.5055\times10^{-6}$ \\
      $\Lambda_6$-$\Lambda_2$   &  $-3.7032$  &  $2.0313$  &  $-1.0038\times10^{-2}$  &  $1.4083\times10^{-2}$  &  $-1.9640\times10^{-3}$  &  $1.1029\times10^{-4}$  &  $-2.1380\times10^{-6}$ \\
      $\Lambda_7$-$\Lambda_2$   &  $-4.8568$  &  $2.3209$  &  $-2.2063\times10^{-2}$  &  $1.8533\times10^{-2}$  &  $-2.5050\times10^{-3}$  &  $1.4020\times10^{-4}$  &  $-2.7305\times10^{-6}$ \\
      $\Lambda_8$-$\Lambda_2$   &  $-8.2442$  &  $2.6203$  &  $-1.8152\times10^{-2}$  &  $2.5720\times10^{-2}$  &  $-3.6087\times10^{-3}$  &  $2.0231\times10^{-4}$  &  $-3.9399\times10^{-6}$ \\
    \hline \hline
  \end{tabular}}
  \label{multipole fitting parameters}
\end{specialtable}

In Fig. \ref{multipole error}, we show the 68\%, 95\%, and 99.7\% relative error of each fit. For each line in the error plot, the corresponding percentage of data points lie below it. We restrict our attention to the domain $1 < \Lambda_2 < 10^4$, as this is the range of $\Lambda_2$ most relevant to current LIGO measurements. The estimate error of each $\Lambda_l$ over this range stays mostly flat with a slight downward trend. (The small ripples that can be seen in the error plots over this range are simply artifacts of how the distribution of NS solutions was computed.) The universality of the multipole relations weaken gradually as $l$ increases, as can be seen in the increasing thickness of the distributions in Fig. \ref{high-order multipoles}. This then increases the maximum estimate error of $\Lambda_l$ for larger $l$ despite the faithfulness of each fit to the shape of the corresponding relation (see Fig. \ref{high-order multipoles}). While 95\% of estimate errors are smaller than $\sim$7\% for the $\Lambda_3$-$\Lambda_2$ relation, 95\% are only smaller than $\sim$50\% for $\Lambda_8$-$\Lambda_2$.

\begin{figure}
	\centering
	\resizebox{0.6\linewidth}{!}{%
	\includegraphics[trim=5 0 5 0, clip]{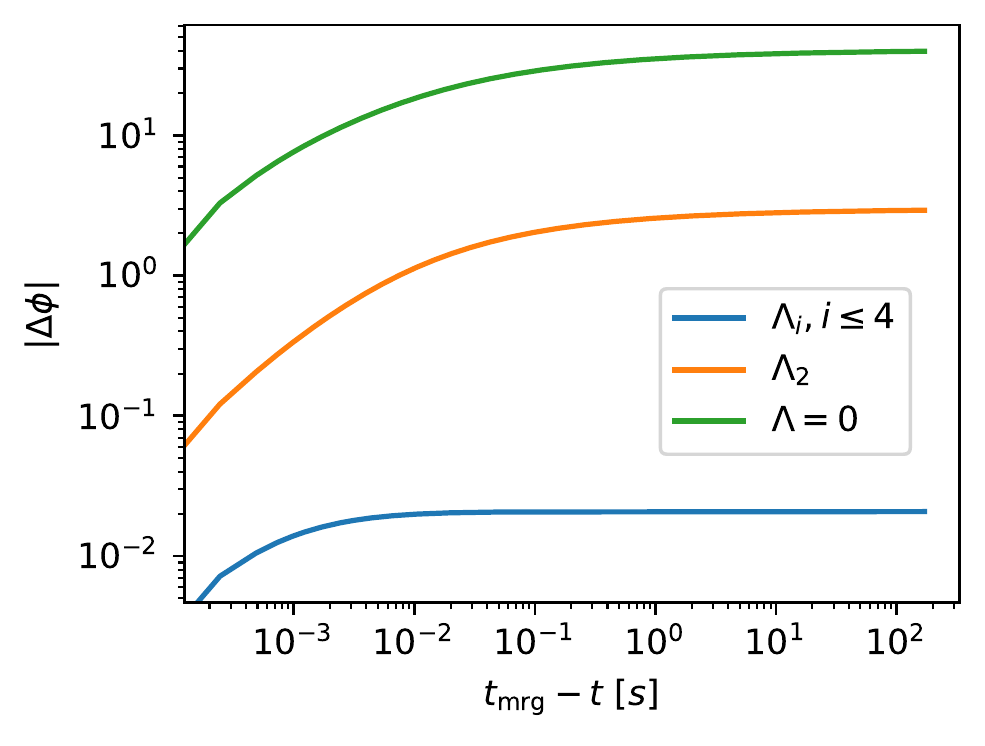}}
	\caption{An example of dephasing between different waveform models (one with no tides, one with only the $l=2$ correction, and one with all corrections up to $l=4$) and the full model (all corrections up to $l=8$). The overall dephasing is very small between the $l\leq4$ model and the full model. Most of the dephasing is accumulated in the last 5 milliseconds, just a few orbits prior to the merger.}
	\label{dephasing}
\end{figure}

The phase of a GW in waveform modelling is affected by the highest order out to which one carries finite-size corrections.\footnote{The leading-terms of the finite-size correction from $\Lambda_l$ is given in \cite{Yagi:2013sva}.} {We demonstrate this with a baseline model of a binary with $m_1 = m_2 = 2.7 M_\odot$ and $\Lambda_1 = \Lambda_2 = 1000$ using the spin-aligned effective-one-body waveform model {\tt TEOBResumS} \cite{Nagar:2018plt}. Often when universal relations are not employed, all finite-size effects are dropped except for the leading-order ($l = 2$) effect. In the baseline model, just the $l=2$ correction alone contributes a phase difference of 36.7 radians compared to a waveform model with no tidal corrections. Further corrections from the $l=3$ and $l=4$ effects using the $\Lambda_3$-$\Lambda_2$ and $\Lambda_4$-$\Lambda_2$ relations respectively incur an additional 2.89 radians. Finally, including the $l = 5$, $6$, $7$, and $8$ corrections using the relations given in this work adds 0.02 radians of dephasing on top of that. (The dephasing between the $l \leq 8$ waveform model and models with fewer corrections is plotted in Fig. \ref{dephasing} as a function of time. For all models, most of the dephasing is accumulated in the last 5 milliseconds before the merger.) Combined, the $l > 2$ corrections contribute 2.91 radians of dephasing.} This demonstrates the importance of the multipole Love relations for faithful waveform modelling.

{The dephasing of the $l > 4$ corrections are currently smaller than GW detector uncertainties, but this could only have been known after fitting to the $l > 4$ multipole relations. Additionally, with the greater sensitivity of future detectors, the $l > 4$ finite-size effects will start to come into view. The order out to which one should carry finite-size corrections in the waveform analysis of actual GW data is dependent on several factors (the EOS model, the signal-to-noise ratio of the merger, etc.); however, in general it is recommended that corrections up to $l = 4$ be included in the analysis of data from current detectors \cite{Yagi:2013sva,Damour:2012yf,Godzieba:2020bbz}.}

\section{\label{R-LT Relation}$R_{M}$-$\tilde{\Lambda}$ Relation}

\begin{figure}[t]
    \centering
    \resizebox{\linewidth}{!}{%
    \subfloat[]{{\includegraphics[trim=25 0 20 0, clip, width=0.45\linewidth]{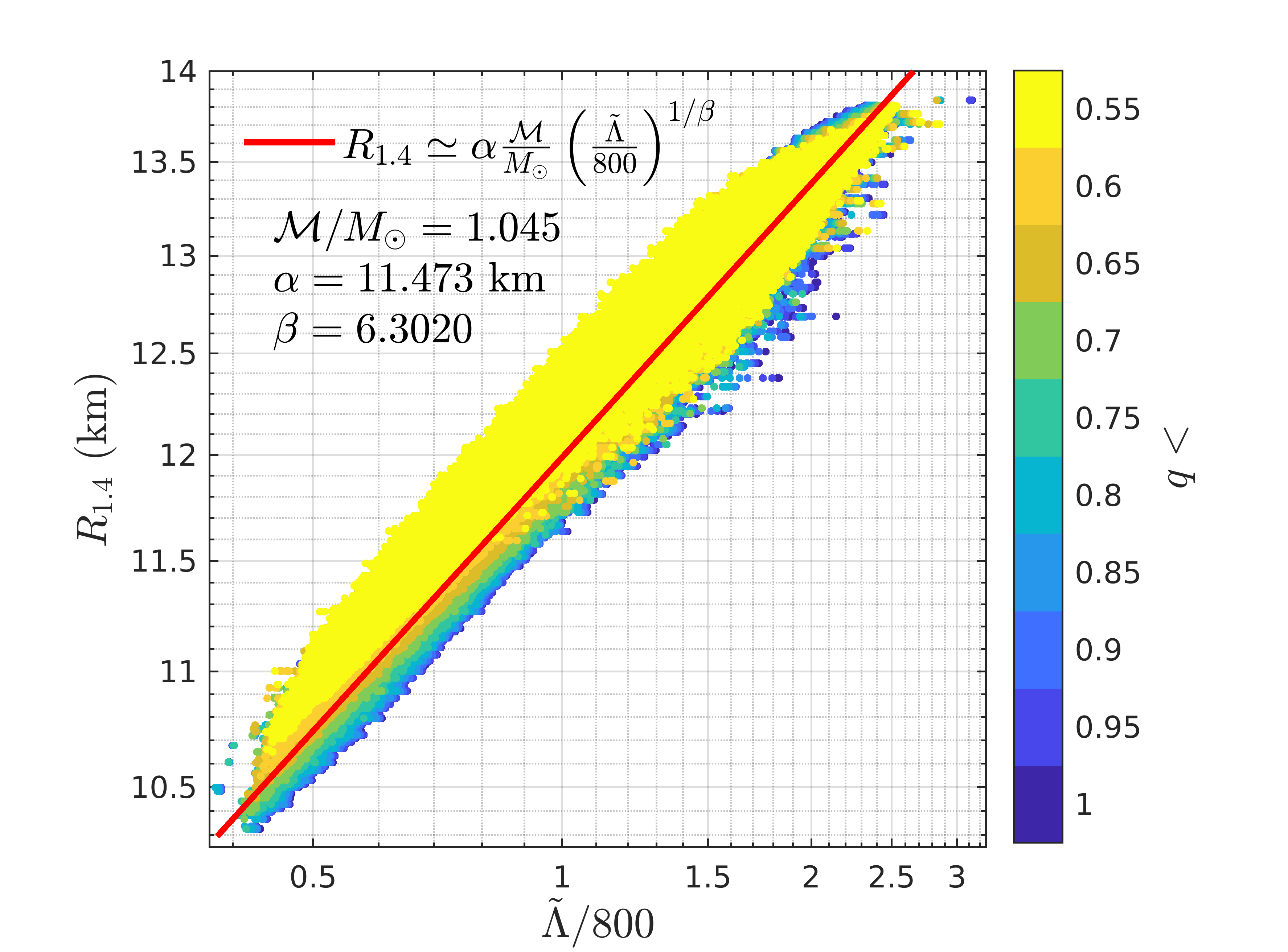}}}%
    \subfloat[]{{\includegraphics[trim=25 0 20 0, clip, width=0.45\linewidth]{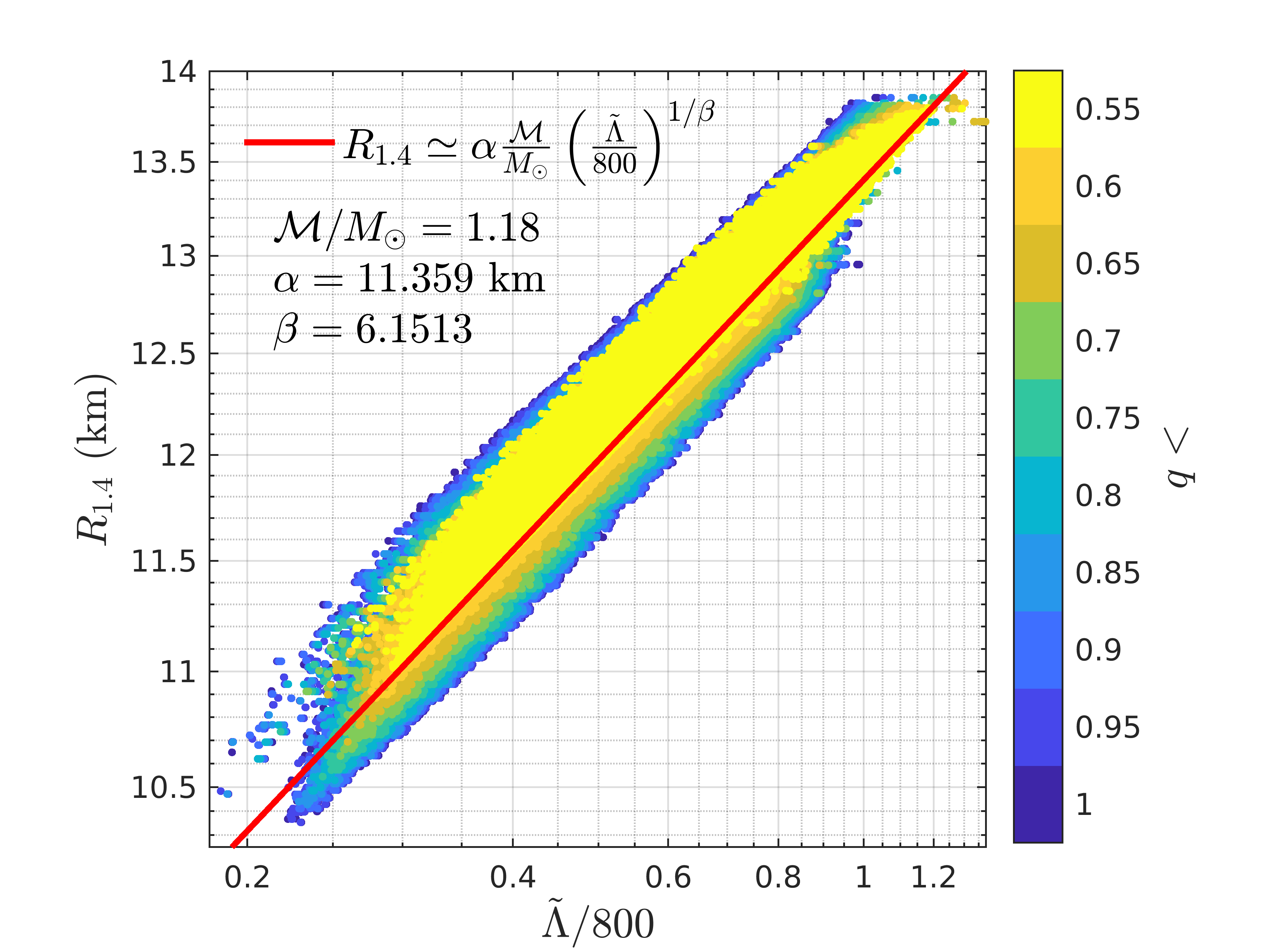}}}}\\
    \resizebox{\linewidth}{!}{%
    \subfloat[]{{\includegraphics[trim=25 0 20 0, clip, width=0.45\linewidth]{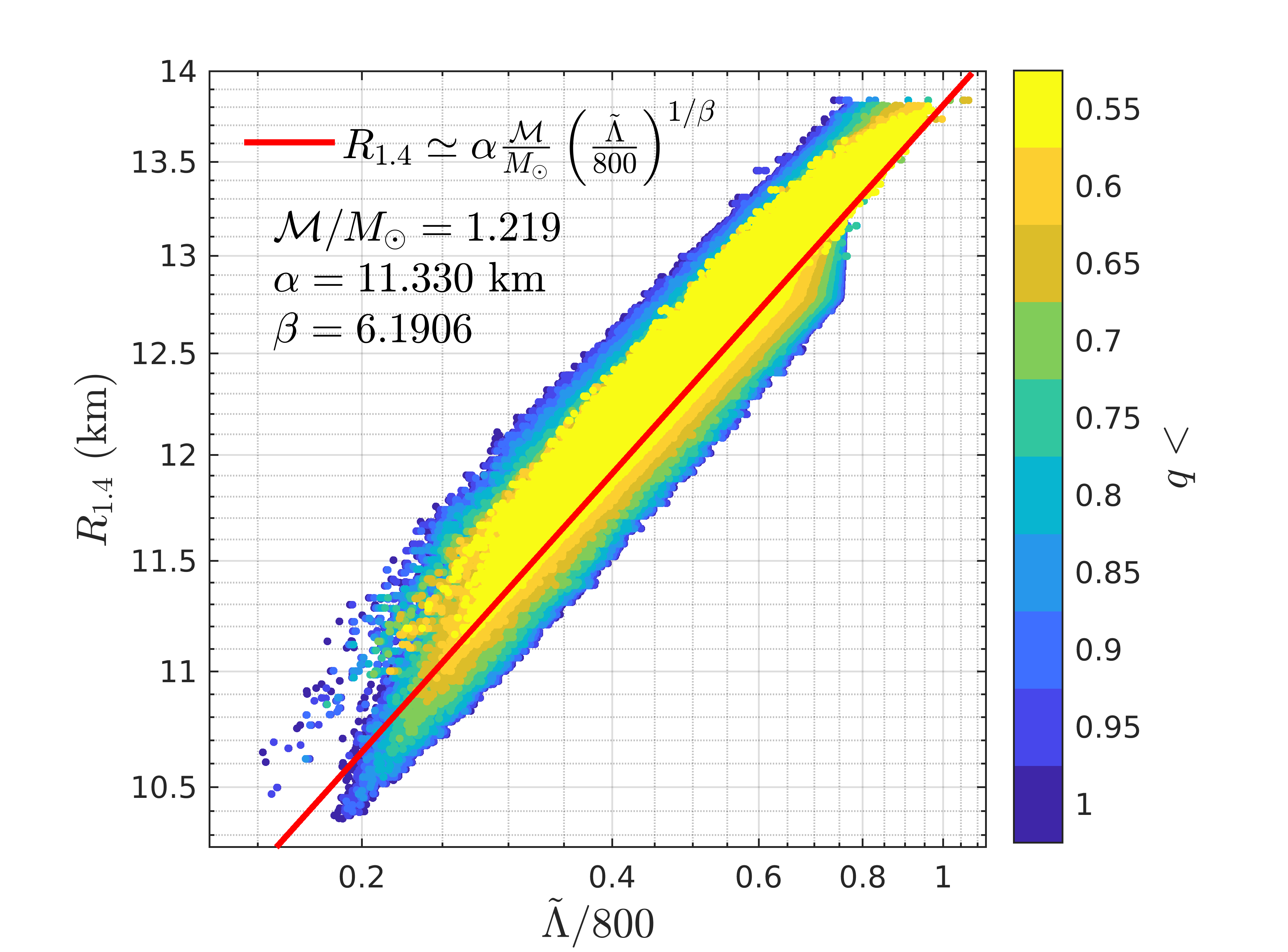}}}%
    \subfloat[]{{\includegraphics[trim=25 0 20 0, clip, width=0.45\linewidth]{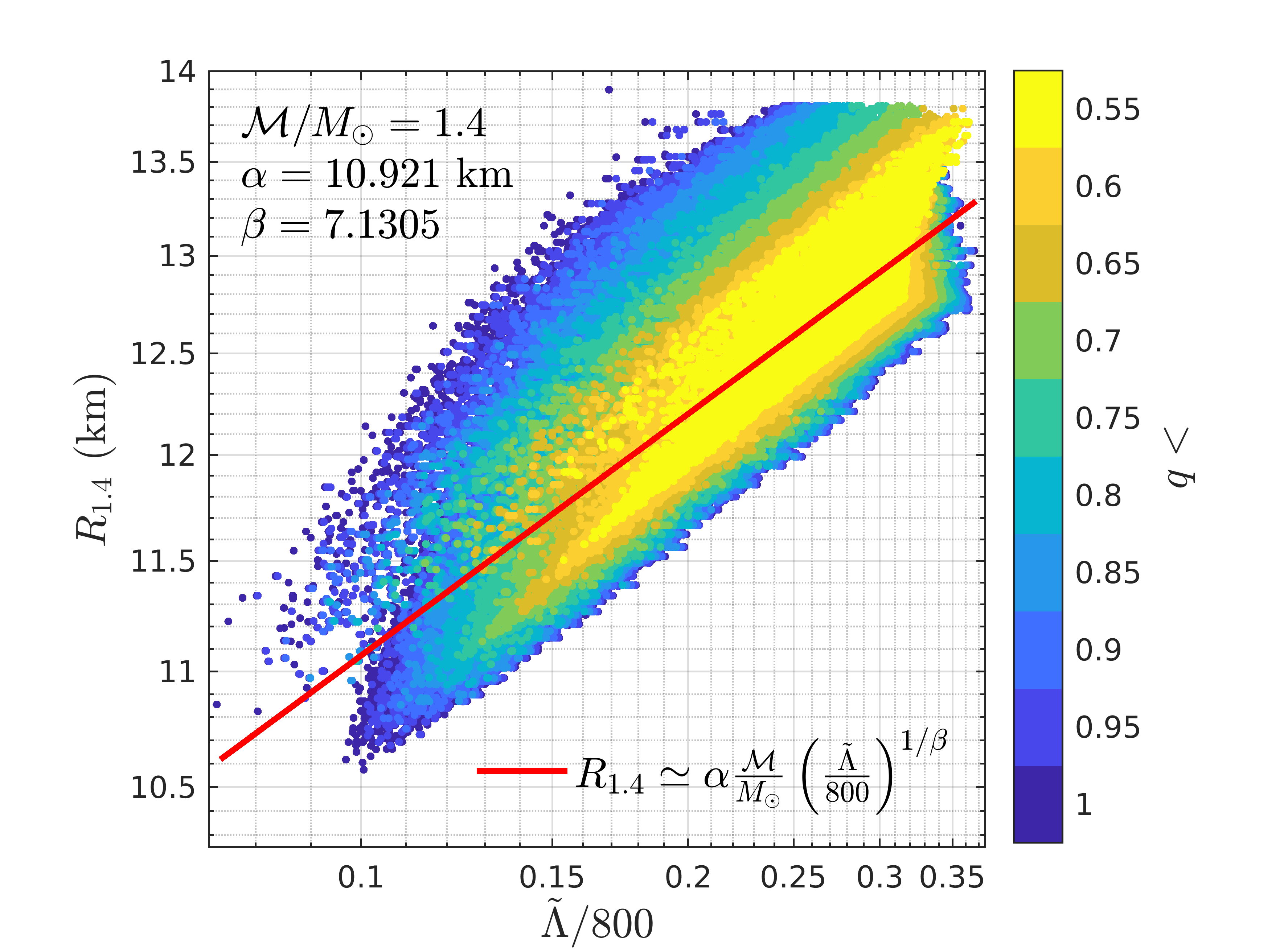}}}}%
    \caption{Example fits to the $R_{1.4}$-$\tilde{\Lambda}$ relation for {(a) $\mathcal{M} = 1.045 M_\odot$, (b) $\mathcal{M} = 1.18 M_\odot$, (c) $\mathcal{M} = 1.219 M_\odot$, and (d) $\mathcal{M} = 1.4 M_\odot$}. Each point represents a BNS and is colored according to the value of the binary mass ratio $q = m_2/m_1$. Points with smaller values of $q$ are drawn on top. The upper limit on the value of $\tilde{\Lambda}$ for each $\mathcal{M}$ derives from the $\Lambda_{2} < 800$ cutoff for the $1.4 M_\odot$ NS imposed on the EOSs generated by our algorithm (see Sec. \ref{Methods}). The relation does not depend on both NSs having the same radius, and indeed for $\mathcal{M}/M_\odot \gtrsim 1.1$ it becomes tighter as $q$ decreases.}%
    \label{R-LT for multiple Mchirp}%
\end{figure}

\begin{figure}[t]
    \centering
    \resizebox{\linewidth}{!}{%
    \subfloat[]{{\includegraphics[trim=25 0 50 0, clip, width=0.5\linewidth]{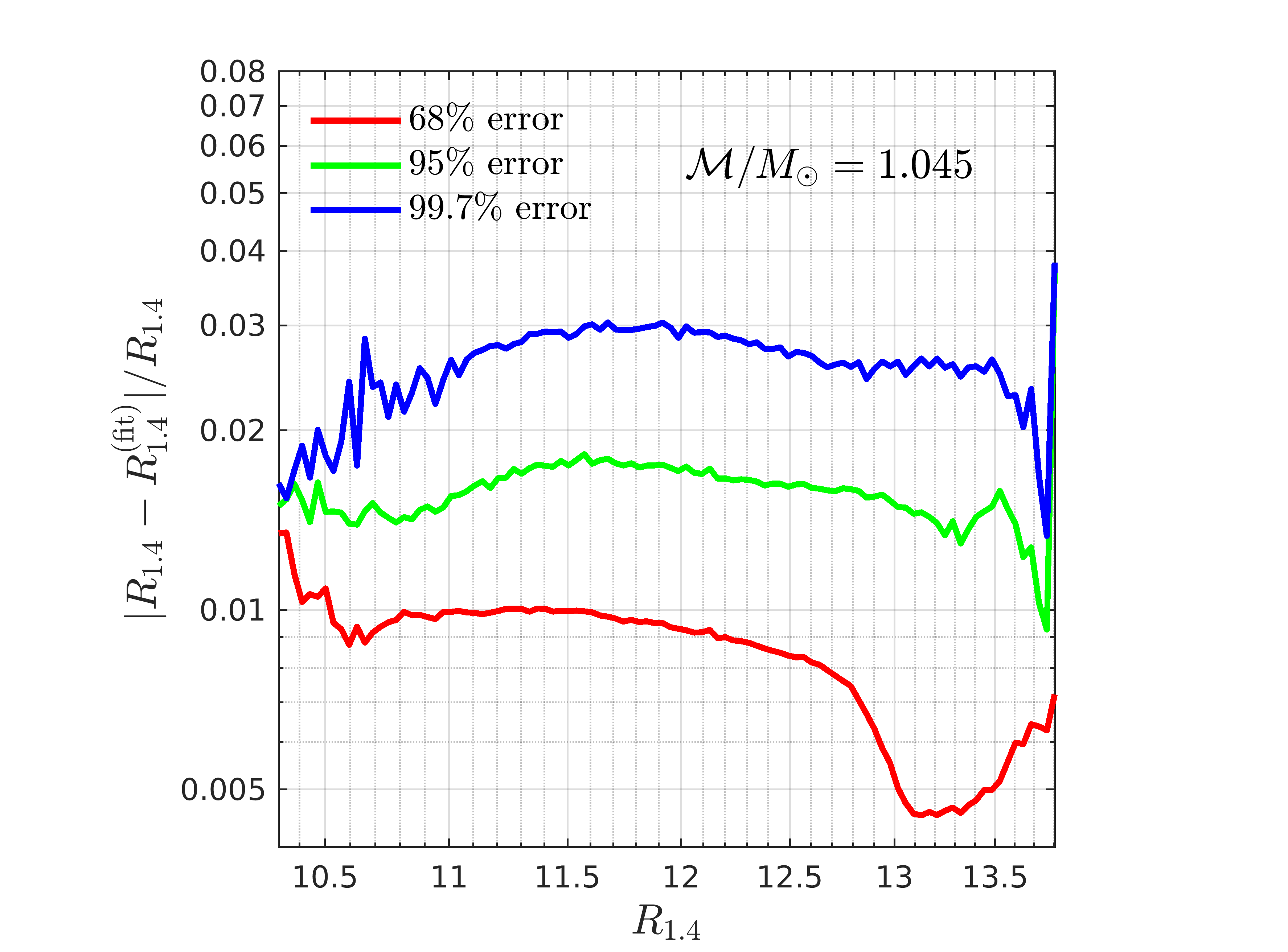}}}%
    \subfloat[]{{\includegraphics[trim=25 0 50 0, clip, width=0.5\linewidth]{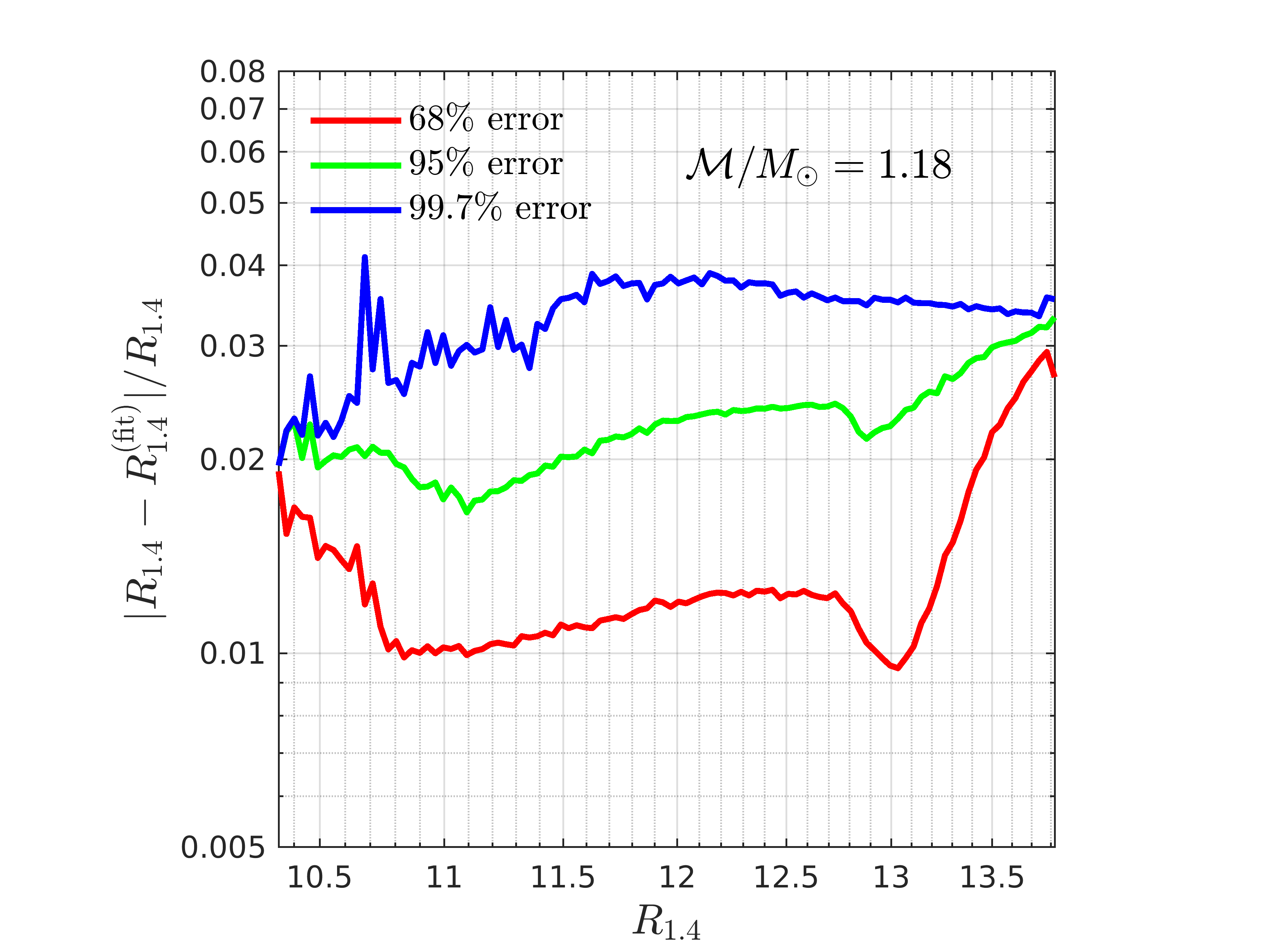}}}}\\
    \resizebox{\linewidth}{!}{%
    \subfloat[]{{\includegraphics[trim=25 0 50 0, clip, width=0.5\linewidth]{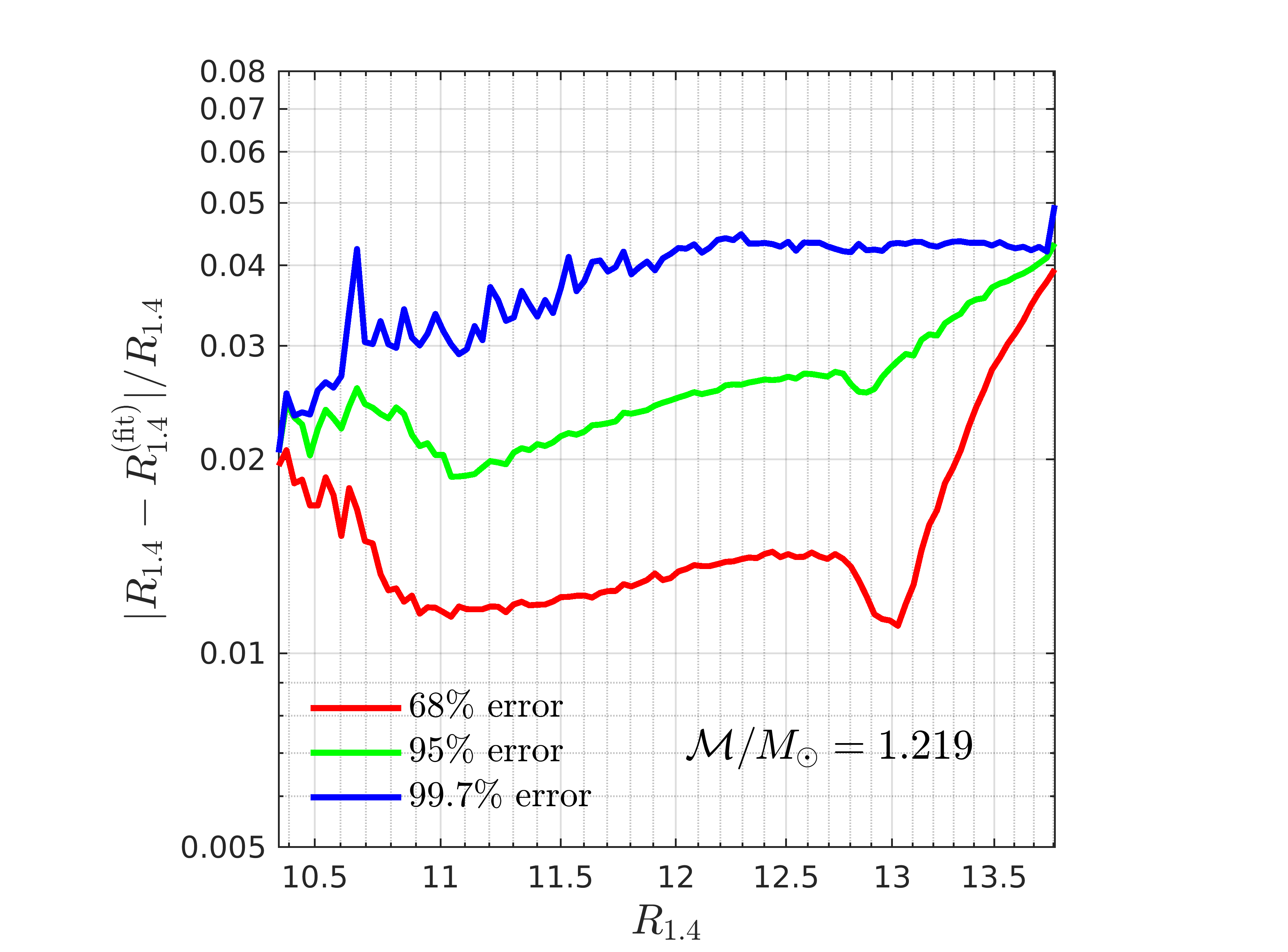}}}%
    \subfloat[]{{\includegraphics[trim=25 0 50 0, clip, width=0.5\linewidth]{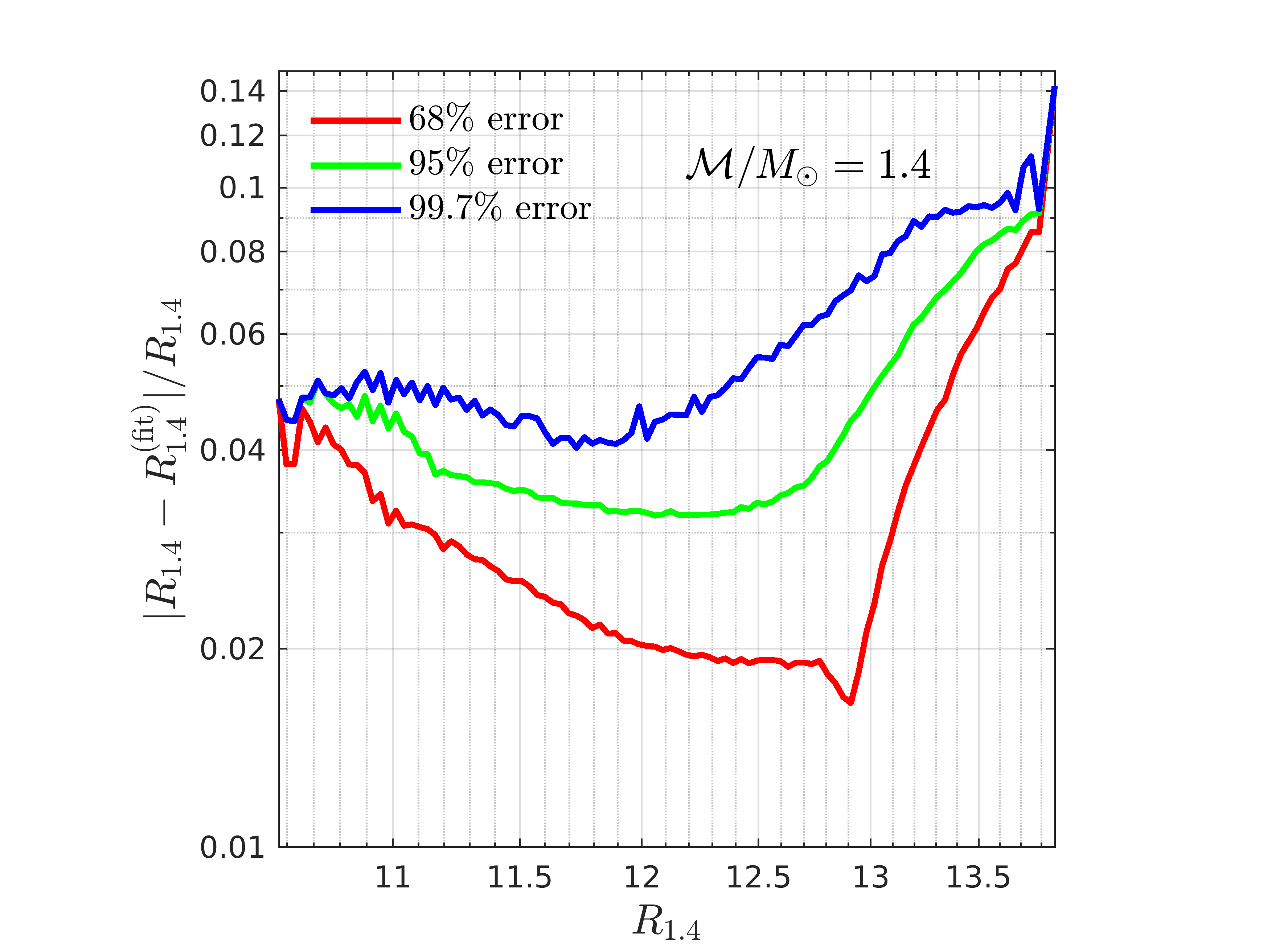}}}}%
    \caption{68\%, 95\%, and 99.7\% relative errors of the fits to the $R_{1.4}$-$\tilde{\Lambda}$ relation for {(a) $\mathcal{M} = 1.045 M_\odot$, (b) $\mathcal{M} = 1.18 M_\odot$, (c) $\mathcal{M} = 1.219 M_\odot$, and (d) $\mathcal{M} = 1.4 M_\odot$}. The error overall stays below $\mathcal{O}(10\%)$, with 95\% of the estimates generally below 4-5\% error, for all values of $\mathcal{M}$.}%
    \label{R-LT error for multiple Mchirp}%
\end{figure}

\begin{figure}
	\centering
	\resizebox{0.6\linewidth}{!}{%
	\includegraphics[trim=50 0 50 0, clip]{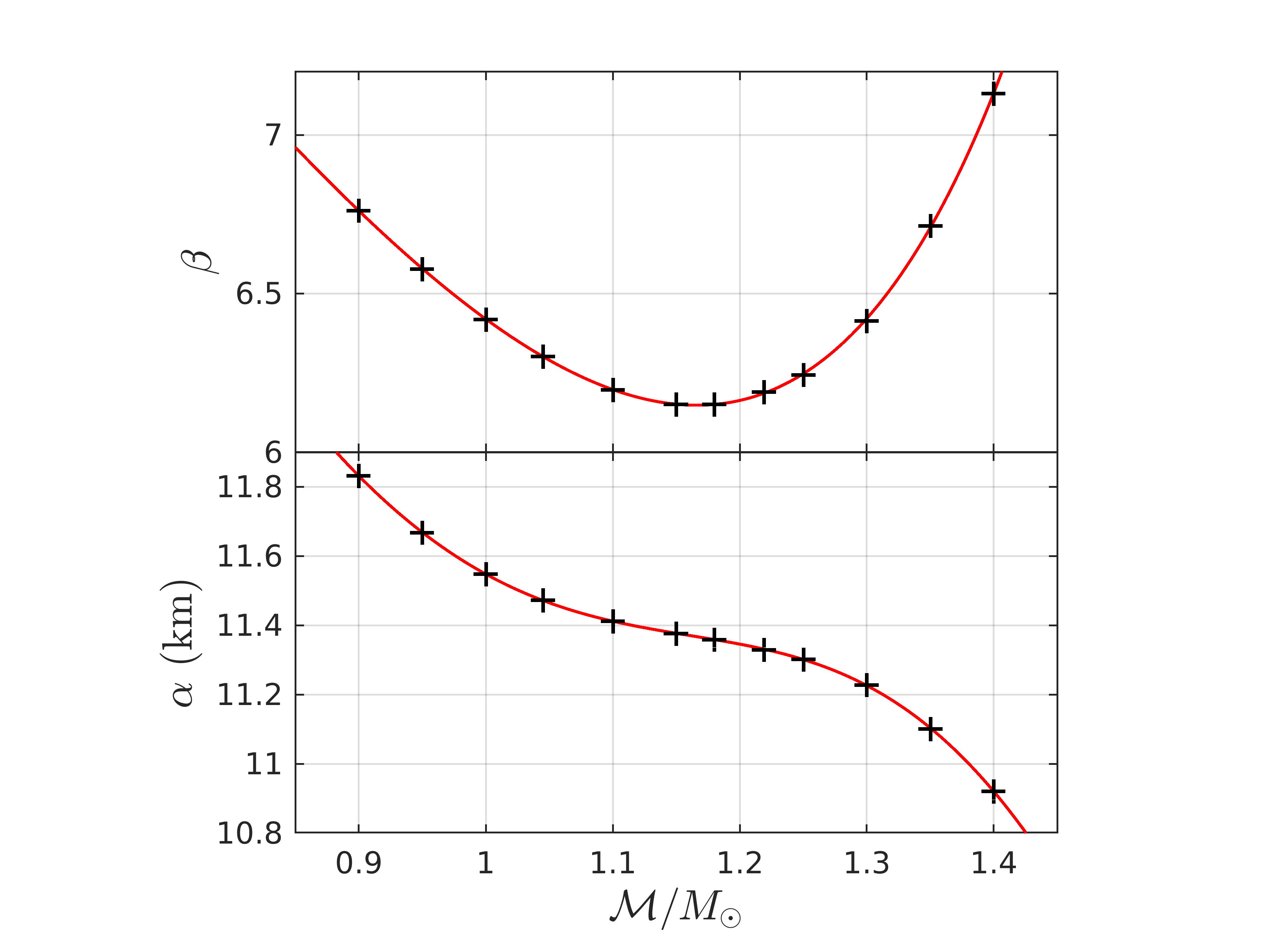}}
	\caption{Fitting parameters $\alpha$ and $\beta$ of the $R_{1.4}$-$\tilde{\Lambda}$ relation as functions of $\mathcal{M}/M_\odot$. $\beta(x)$ does not vary monotonically with $x$, but has a minimum.}
	\label{chirp mass variation}
\end{figure}

\begin{figure}
	\centering
	\resizebox{0.6\linewidth}{!}{%
	\includegraphics[trim=40 0 50 0, clip]{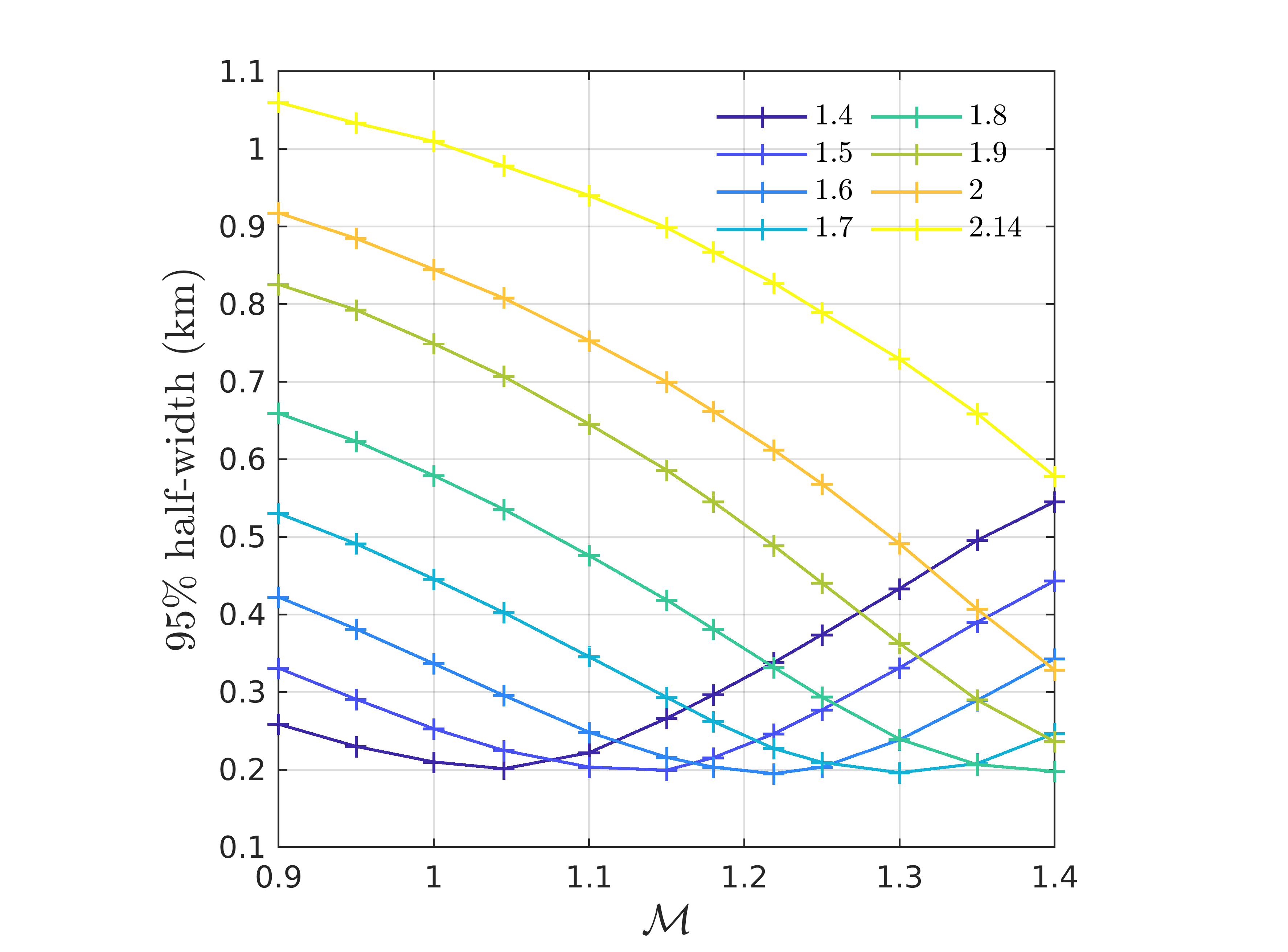}}
	\caption{The approximate uncertainty of the estimated value of $R_{M}$ computed using the $R_{M}$-$\tilde{\Lambda}$ relation as a function of $\mathcal{M}$. Each curve is colored according to the value of $M$ (given in units of $M_\odot$). The uncertainty is defined as the half-width of the symmetric interval centered at $\Delta R_M = R_M - R_M^{\rm (fit)} = 0$ that encloses 95\% of the data points in the histogram of $\Delta R_M$. For every $M$, there is an optimal value of $\mathcal{M}$ such that this uncertainty is minimized.}
	\label{half-width plot}
\end{figure}

We analyze the $R_{1.4}$-$\tilde{\Lambda}$ relation at twelve different fixed values of the chirp mass $\mathcal{M}$, which are given in Table \ref{R-LT fitting parameters}. We compute between 750,000 and 1,000,000 valid individual binaries for each value of $\mathcal{M}$. Several example plots of the relation are shown in Fig. \ref{R-LT for multiple Mchirp}. The relation's dependence on the binary mass ratio $q$ is illustrated by the coloring of the points in these plots. Each point in the plot represents a BNS. Points with smaller values of $q$ are plotted on top. An important conclusion to draw from these plots is that the relation does not depend upon both stars having the same radius \cite{Zhao:2018nyf}. For $\mathcal{M} \lesssim 1.25 M_\odot$, the relation remains fairly tight for all values of $q$. Further, for $\mathcal{M} \gtrsim 1.1 M_\odot$\footnote{The smallest physical value $\mathcal{M}$ can take is when $m_1 = m_2 = M_{\rm min} \approx 1.15 M_\odot$ (see Sec. \ref{Methods}). Using Eq. (\ref{Mchirp definition}), this gives us $\mathcal{M} \gtrsim 1.001 M_\odot$. Since we permit $m_1$ and $m_2$ to be less than $1.15 M_\odot$, we are able to reach as low as $\mathcal{M} = 0.9 M_\odot$.}, the relation actually becomes \textit{tighter} as $q$ decreases (i.e. as the radii of the two stars differ more and more), which can be understood by considering the definition of $\tilde{\Lambda}$ in Eq. (\ref{tildeLambda}). For fixed $R_{1.4}$, the range of possible values $\tilde{\Lambda}$ can take is constrained by $q$. When $q = 1$, the masses of the binary can span the range from the minimum to the maximum mass, $M_{\rm min} \leq m_1 = m_2 \leq M_{\rm max}$. Hence, $\min{(\Lambda_{2})} \leq \Lambda_{2,1} = \Lambda_{2,2} \leq \max{(\Lambda_{2})}$, and $\min{(\Lambda_2)} \leq \tilde{\Lambda} \leq \max{(\Lambda_2)}$. As $q$ decreases, the bounds for both $m_1$ and $m_2$ shrink and no longer overlap, causing the same to happen for $\Lambda_{2,1}$ and $\Lambda_{2,2}$. This, as we see, also shrinks the bounds on $\tilde{\Lambda}$. Thus, we expect the relation to tighten as $q$ decreases.

We construct a fitting function for the $R_{1.4}$-$\tilde{\Lambda}$ relation by considering a slightly generalized form of Eq. (\ref{R-LT relation}): \begin{equation}
    R_{M} \simeq \alpha \frac{\mathcal{M}}{M_\odot} \left( \frac{\tilde{\Lambda}}{800} \right)^{1/\beta}, \label{general R-LT relation}
\end{equation} where, in this case, $M = 1.4 M_\odot$. Here the proportionality constant $\alpha$ and the inverse exponent $\beta$ are the fitting parameters and, consequently, will be dependent on $\mathcal{M}$. These fits are also shown in Fig. \ref{R-LT for multiple Mchirp}. The fitting parameters for all values of $\mathcal{M}$ are given in Table \ref{R-LT fitting parameters}. A sense of the accuracy of the estimated value of $R_{1.4}$ from the fit can be gathered from Fig. \ref{R-LT error for multiple Mchirp}, where we plot the 68\%, 95\%, and 99.7\% relative error of the example fits. Overall, the estimates are accurate to within $\mathcal{O}(10\%)$ error for all values of $\mathcal{M}$, and are, in fact, accurate to within $\sim$5\% for most values of $R_{1.4}$.

We can extend our fitting results to all $\mathcal{M} \in [0.9 M_\odot, 1.4 M_\odot]$ by fitting the dependence of $\alpha$ and $\beta$ on $\mathcal{M}$. We construct the rational fitting functions \begin{equation}
    \alpha(x) = \frac{\sum_{k=0}^{3} p_k x^k}{\sum_{k=0}^{2} q_k x^k} \ {\rm km} \qquad {\rm and} \qquad  \beta(x) = \frac{\sum_{k=0}^{2} p_k x^k}{\sum_{k=0}^{2} q_k x^k}, \label{alpha and beta}
\end{equation} where $x = (\mathcal{M}/M_\odot - \mu_\mathcal{M})/\sigma_\mathcal{M}$, $\mu_\mathcal{M} = 1.1537$, and $\sigma_\mathcal{M} = 0.15927$. These fits, which are in excellent agreement the values in Table \ref{R-LT fitting parameters}, are shown in Fig. \ref{chirp mass variation}. The fitting parameters $\vec{p} = \{p_k\}$ and $\vec{q} = \{q_k\}$ for $\alpha(x)$ and $\beta(x)$ are given in Table \ref{alpha and beta fitting parameters}. What is interesting is that the inverse exponent $\beta$ is not monotonic. Rather, it has a minimum at $\mathcal{M} = 1.1661 M_\odot$. A possible contributor to this effect is the decrease in the variety of possible binaries as $\mathcal{M}$ increases. The maximum value $\mathcal{M}$ could take for a given EOS is found by letting $m_1 = m_2 = M_{\rm max}$ in Eq. (\ref{Mchirp definition}), which yields $\mathcal{M}_{\rm max} = 2^{-1/5} M_{\rm max}$. For $1/2 \leq q \leq 1$, $m_1$ and $m_2$ are bounded by \begin{align}
    2^{1/5}\mathcal{M} &\leq m_1 \leq \min{(12^{1/5}\mathcal{M},M_{\rm max})}, \label{m1 bounds}\\
    (3/8)^{1/5} \mathcal{M} &\leq m_2 \leq 2^{1/5} \mathcal{M}. \label{m2 bounds}
\end{align} Hence, as $\mathcal{M}$ increases, the relation becomes gradually dominated by (1) EOSs with $M_{\rm max} \geq 2^{1/5} \mathcal{M}$ and (2) only those BNSs from each EOS that lie in the increasingly narrow range in Eq. (\ref{m1 bounds}). This decrease of BNS variety could play a role in the non-monotonic behavior of $\beta(x)$.

One could, of course, consider more generally the relation between $\tilde{\Lambda}$ and the radius of a NS with some mass $M$, $R_{M}$. We pursue this thought by looking at $R_M$-$\tilde{\Lambda}$ for $M/M_\odot = 1.4$, $1.5$, $1.6$, $1.7$, $1.8$, $1.9$, $2$, and $2.14$. Just as for the $R_{1.4}$-$\tilde{\Lambda}$ relation, we utilize the fitting function in Eq. (\ref{R-LT relation}), and then find $\alpha$ and $\beta$ as functions of $x$ using Eq. (\ref{alpha and beta}). In Table \ref{alpha and beta fitting parameters}, we show the fitting parameters $\vec{p} = \{p_k\}$ and $\vec{q} = \{q_k\}$ of $\alpha(x)$ and $\beta(x)$ for each $M$. The tightness of the $R_{M}$-$\tilde{\Lambda}$ relation (and thus the general quality of the estimate from the fit) is dependent on both $M$ and $\mathcal{M}$. We illustrate this in Fig. \ref{half-width plot} by plotting the approximate uncertainty of the estimated value of $R_M$ as a function of $\mathcal{M}$ for several values of $M$. Since our EOSs and BNSs do not come from prior probability distributions, we non-stringently define the uncertainty here as the half-width of the symmetric interval centered at $\Delta R_M = R_M - R_M^{\rm (fit)} = 0$ that encloses 95\% of the data points in the histogram of $\Delta R_M$ for fixed $\mathcal{M}$. Interestingly, the uncertainty for each $M$ reaches a minimum at some particular value of $\mathcal{M}$, with the minimum uncertainty for each $M$ being around 0.2 km in the range of $\mathcal{M}$ we considered. The minima for $M = 1.4 M_\odot$ through $1.8 M_\odot$ are visible in the Fig. \ref{half-width plot}. This reveals that there is an optimal $M$ for each $\mathcal{M}$ such that $R_M$ is maximally constrained by the $R_M$-$\tilde{\Lambda}$ relation at that $\mathcal{M}$. Thus, for example, a chirp mass of $\mathcal{M} \approx 1.05 M_\odot$ would yield the best estimate of $R_{1.4}$, while a chirp mass of $\mathcal{M} \approx 1.4 M_\odot$ would yield the best estimate of $R_{1.8}$. Further, there appears to be a linear dependence of the optimal $M$ on $\mathcal{M}$; however, a wider range of $\mathcal{M}$ would need to be considered to confirm this. The change in the variety of BNSs as $\mathcal{M}$ increases, as previously described, may contribute to this. At larger $\mathcal{M}$, the relation becomes dominated by larger mass NSs; thus, the relation may become more sensitive to the radii of larger mass NSs as $\mathcal{M}$ increases.

The $R_M$-$\tilde{\Lambda}$ relation, then, allows one to use any binary to place a robust, EOS-agnostic constraint on $R_M$ using just $\tilde{\Lambda}$ and $\mathcal{M}$. This offers great prospects for the upcoming LIGO O4 run and for third-generation detectors. The O4 run expects to see $10^{+52}_{-10}$ detections within a search volume of $1.6\times10^7$ Mpc$^3$ year \cite{KAGRA:2013rdx}. Every BNS detection can be transformed into a maximum constrain on some $R_M$. However, even the weaker constraints afforded by $R_M$-$\tilde{\Lambda}$ are of still great utility. Just 10 weak constraints on $R_{1.4}$ using the $R_{1.4}$-$\tilde{\Lambda}$ relation will yield a reliable value for $R_{1.4}$. Further, a reduction in statistical uncertainty thanks to increased sensitivity improves the effectiveness of universal relations, as, for example, the systematic errors of fits to multipole relations are generally smaller than statistical uncertainty \cite{Yagi:2013sva}.

\begin{specialtable}[t]
  \caption{Fitting parameters of the general $R_{1.4}$-$\tilde{\Lambda}$ relation given in Eqs. (\ref{general R-LT relation}) for different values of $\mathcal{M}$.}
  \centering
  \begin{tabular}{ccc}
    \hline \hline
      $\mathcal{M}/M_\odot$  & $\alpha$ (km) & $\beta$ \\
    \hline
      0.900  &  11.832  &  6.7621  \\
      0.950  &  11.668  &  6.5775  \\
      1.000  &  11.548  &  6.4189  \\
      1.045  &  11.473  &  6.3020  \\
      1.100  &  11.412  &  6.1972  \\
      1.150  &  11.377  &  6.1515  \\
      1.180  &  11.359  &  6.1513  \\
      1.219  &  11.330  &  6.1906  \\
      1.250  &  11.302  &  6.2441  \\
      1.300  &  11.228  &  6.4147  \\
      1.350  &  11.102  &  6.7139  \\
      1.400  &  10.921  &  7.1305  \\
    \hline \hline
  \end{tabular}
  \label{R-LT fitting parameters}
\end{specialtable}

\begin{specialtable}[t]
  \caption{Fitting parameters $\vec{p} = \{p_k\}$ and $\vec{q} = \{q_k\}$ for $\alpha$ and $\beta$ as functions of $x = (\mathcal{M}/M_\odot - \mu_\mathcal{M})/\sigma_\mathcal{M}$, where $\mu_\mathcal{M} = 1.1537$  and $\sigma_\mathcal{M} = 0.15927$ for several values of $M$. The fitting functions are given in Eq. (\ref{alpha and beta fitting parameters}).}
  \centering
  \begin{tabular}{ccccccccc}
    \hline \hline
     $M/M_\odot$ &    & $p_{0}$ & $p_{1}$ & $p_{2}$ & $p_{3}$ & $q_{0}$ & $q_{1}$ & $q_{2}$ \\
    \hline
      \multirow{2}{*}{1.4}  & $\beta(x)$ &  $404.40$  &  $-96.991$  &  $26.475$  &  -  &  $65.755$  &  $-15.259$  &  $1$  \\
       &  $\alpha(x)$  &  $224.75$  &  $-24.553$  &  $11.832$  &  $-1.8434$  &  $19.758$  &  $-1.9914$  &  $1$  \\
    \hline
    \multirow{2}{*}{1.5}  & $\beta(x)$ &  $502.01$  &  $-119.44$  &  $32.193$  &  -  &  $79.153$  &  $-16.598$ &  $1$  \\
         & $\alpha(x)$  &  $282.86$  &  $-29.568$  &  $12.893$  &  $-2.2628$  &  $24.833$  &  $-2.3357$  &  $1$  \\
    \hline
    \multirow{2}{*}{1.6}  & $\beta(x)$ &  $642.10$  &  $-152.88$  &  $40.447$  &  -  &  $98.054$  &  $-18.391$  &  $1$  \\
         & $\alpha(x)$  &  $386.63$  &  $-42.102$  &  $14.780$  &  $-3.0054$  &  $33.942$  &  $-3.2743$  &  $1$  \\
    \hline
    \multirow{2}{*}{1.7}  & $\beta(x)$ &  $877.56$  &  $-210.98$  &  $54.468$  &  -  &  $129.67$  &  $-21.419$ &  $1$  \\
         & $\alpha(x)$  &  $598.97$  &  $-73.554$  &  $18.854$  &  $-4.5818$  &  $52.655$  &  $-5.7193$  &  $1$  \\
    \hline
    \multirow{2}{*}{1.8}  & $\beta(x)$ &  $1442.2$  &  $-356.28$  &  $88.775$  &  -  &  $206.20$  &  $-29.608$  &  $1$  \\
         & $\alpha(x)$  &  $1291.3$  &  $-192.50$  &  $32.831$  &  $-10.064$  &  $113.87$  &  $-15.207$  &  $1$  \\
    \hline
    \multirow{2}{*}{1.9}  & $\beta(x)$ &  $2734.1$  &  $-709.62$  &  $174.36$  &  -  &  $380.72$  &  $-46.815$ &  $1$  \\
         & $\alpha(x)$  &  $3282.2$  &  $-592.78$  &  $80.705$  &  $-27.980$  &  $291.00$  &  $-48.054$  &  $1$  \\
    \hline
    \multirow{2}{*}{2}  & $\beta(x)$ &  $63061$  &  $-19071$  &  $3663.7$  &  -  &  $8959.4$  &  $-1478.5$ &  $1$  \\
         & $\alpha(x)$  &  $9986.9$  &  $-2011.9$  &  $232.61$  &  $-82.091$  &  $887.58$  &  $-167.09$  &  $1$  \\
    \hline
    \multirow{2}{*}{2.14}  & $\beta(x)$ &  $93209$  &  $-34415$  &  $6102.8$  &  -  &  $12686$  &  $-2286.4$ &  $1$  \\
         & $\alpha(x)$  &  $58353$  &  $-18272$  &  $1904.6$  &  $-609.58$  &  $5234.8$  &  $-1543.2$  &  $1$  \\
    \hline \hline
  \end{tabular}
  \label{alpha and beta fitting parameters}
\end{specialtable}

\section{\label{Conclusions}Conclusions}

We supplement the tool set of GW analysis and waveform modelling by presenting entirely new fits to several universal relations between high-multipole-order dimensionless gravitoelectric tidal deformabilities $\Lambda_l$ and to the universal relation for BNS between the radius of the $1.4 M_\odot$ NS, $R_{1.4}$, and the reduced tidal deformability $\tilde{\Lambda}$. We compute these utilizing a data set of nearly two-million phenomenological EOS sampled from across a broad parameter space using an MCMC algorithm.

First, we present fits to the multipole relations. Previous fits \cite{Yagi:2013sva,Godzieba:2020bbz} had been made to just the $\Lambda_3$-$\Lambda_2$ and $\Lambda_4$-$\Lambda_2$ relations. We extend the library of fits by looking at the $\Lambda_5$-$\Lambda_2$, $\Lambda_6$-$\Lambda_2$, $\Lambda_7$-$\Lambda_2$, and $\Lambda_8$-$\Lambda_2$ relations. The tightness of the relations weakens as $l$ increases. Consequently, though the fits are faithful to the shapes of the relations, the maximum estimate error of the fits increases to the order of 50\% for $\Lambda_8$. {The inclusion of the finite-size effects of the $l < 4$ multipoles in waveform analysis can incur as much as 0.02 radians of dephasing compared to including only the $l \leq 4$ effects. Collectively, the $l > 2$ effects contribute as much as 2.91 radians of dephasing, and it is recommended that finite-size corrections for $l > 2$ multipoles be included in the analysis of GW data wherever they are at least comparable to detector uncertainties.} The {full} usefulness of these $l > 4$ relations in GW data analysis will be realized with the increased sensitivity of the upcoming third-generation GW detectors like LIGO III \cite{Adhikari:2013kya}, the Einstein Telescope \cite{ET-GW,Hild:2008ng}, and Cosmic Explorer \cite{Reitze:2019iox}, as the finite-size effects of these multipole orders are currently smaller than measurement error.

Next, we analyze the $R_{1.4}$-$\tilde{\Lambda}$ relation. The original derivation of the relation \cite{Zhao:2018nyf} yields an expression, given in Eq. (\ref{R-LT relation}), that is linearly dependent on the chirp mass $\mathcal{M}$ of the BNS. Fitting the relation for different fixed values of $\mathcal{M}$ reveals any nonlinear dependence the relation may have on $\mathcal{M}$ and allows us to compute an expression that more accurately estimates $R_{1.4}$. We do this for twelve different values of $\mathcal{M}$ between $0.9 M_\odot$ and $1.4 M_\odot$, using the fitting function in Eq. (\ref{general R-LT relation}). We then interpolate the fitting parameters to all $\mathcal{M} \in [0.9 M_\odot, 1.4 M_\odot]$ by fitting them as functions of $\mathcal{M}$. The accuracy of the estimate of $R_{1.4}$ for any value of $\mathcal{M}$ is found to be quite good. 95\% of the estimates are within $\sim$5\% of $R_{1.4}$.

We then consider a generalized form of the relation $R_{M}$-$\tilde{\Lambda}$ for a generic NS mass $M$. We perform the same analysis as for the $R_{1.4}$-$\tilde{\Lambda}$ relation for seven other values of $M$. We find that the level of uncertainty in the estimate of $R_M$ depends on both $M$ and $\mathcal{M}$. There is, in fact, an optimal value of $M$ for each $\mathcal{M}$ such that $R_M$ is maximally constrained by the relation at that value of $\mathcal{M}$. Therefore, this relation will be an excellent tool for combining the results from multiple GW detections of BNSs into constraints on NS radii.

The parameter space of possible EOS explored by our MCMC algorithm to compute our EOS data set can be further restricted with the inclusion of possible future LIGO/Virgo/KAGRA O4 constraints, laboratory constraints, such as those from heavy-ion collisions \cite{Danielewicz:2002pu} and PREX \cite{Reed:2021nqk}, X-ray burst observations from NICER \cite{Steiner:2010fz,Lattimer:2012nd,Ozel:2016oaf}, and by combining the phenomenological EOSs with results from pQCD calculations \cite{Annala:2019puf}.


\vspace{6pt} 



\authorcontributions{D.G. generated and analyzed the TOV data, D.G. and D.R. interpreted the results and wrote the manuscript. All authors have read and agreed to the published version of the manuscript.}

\funding{This research was funded by U.S. Department of Energy, Office of Science, Division of Nuclear Physics under Award Number(s) DE-SC0021177 and by the National Science Foundation under Grants No. PHY-2011725 and PHY-2116686. Computations for this research were performed on the Pennsylvania State University's Institute for Computational and Data Sciences Advanced CyberInfrastructure (ICDS-ACI).}

\dataavailability{EOS parameters and bulk properties of reference NSs generated for this work are publicly available on Zenodo \citep{godzieba_daniel_a_2020_3954899}.} 

\acknowledgments{The authors gratefully wish to acknowledge R. Gamba for waveform dephasing calculations and S. Bernuzzi for discussions that motivated us to start this work. The authors also wish to thank R. Gamba for discovering errors in the previous version of the manuscript.}

\conflictsofinterest{The authors declare no conflict of interest. The funders had no role in the design of the study; in the collection, analyses, or interpretation of data; in the writing of the manuscript, or in the decision to publish the~results.} 



\abbreviations{The following abbreviations are used in this manuscript:\\

\noindent 
\begin{tabular}{@{}ll}
EOS & Equation of state\\
NS & Neutron star\\
BNS & Binary neutron star\\
GW & Gravitational wave\\
MCMC & Markov chain Monte Carlo\\
TOV & Tolman-Oppenheimer-Volkoff
\end{tabular}}




\end{paracol}
\reftitle{References}


\externalbibliography{yes}
\bibliography{biblio}

\begin{thebibliography}{999}

\bibitem[{Andersson} and {Kokkotas}(1996)]{Andersson:1996pn}
{Andersson}, N.; {Kokkotas}, K.D.
\newblock {Gravitational Waves and Pulsating Stars: What Can We Learn from
  Future Observations?}
\newblock {\em Phys. Rev. Lett.} {\bf 1996}, {\em 77},~4134--4137,
  \href{http://xxx.lanl.gov/abs/gr-qc/9610035}{{\normalfont
  [arXiv:gr-qc/gr-qc/9610035]}}.
\newblock
  doi:{\changeurlcolor{black}\href{https://doi.org/10.1103/PhysRevLett.77.4134}{\detokenize{10.1103/PhysRevLett.77.4134}}}.

\bibitem[{Andersson} and {Kokkotas}(1998)]{Andersson:1997rn}
{Andersson}, N.; {Kokkotas}, K.D.
\newblock {Towards gravitational wave asteroseismology}.
\newblock {\em Mon. Not. Roy. Astron. Soc.} {\bf 1998}, {\em 299},~1059--1068,
  \href{http://xxx.lanl.gov/abs/gr-qc/9711088}{{\normalfont
  [arXiv:gr-qc/gr-qc/9711088]}}.
\newblock
  doi:{\changeurlcolor{black}\href{https://doi.org/10.1046/j.1365-8711.1998.01840.x}{\detokenize{10.1046/j.1365-8711.1998.01840.x}}}.

\bibitem[{Benhar} \em{et~al.}(1999){Benhar}, {Berti}, and
  {Ferrari}]{Benhar:1998au}
{Benhar}, O.; {Berti}, E.; {Ferrari}, V.
\newblock {The imprint of the equation of state on the axial w-modes of
  oscillating neutron stars}.
\newblock {\em Mon. Not. Roy. Astron. Soc.} {\bf 1999}, {\em 310},~797--803,
  \href{http://xxx.lanl.gov/abs/gr-qc/9901037}{{\normalfont
  [arXiv:gr-qc/gr-qc/9901037]}}.
\newblock
  doi:{\changeurlcolor{black}\href{https://doi.org/10.1046/j.1365-8711.1999.02983.x}{\detokenize{10.1046/j.1365-8711.1999.02983.x}}}.

\bibitem[{Benhar} \em{et~al.}(2004){Benhar}, {Ferrari}, and
  {Gualtieri}]{Benhar:2004xg}
{Benhar}, O.; {Ferrari}, V.; {Gualtieri}, L.
\newblock {Gravitational wave asteroseismology reexamined}.
\newblock {\em Phys. Rev. D} {\bf 2004}, {\em 70},~124015,
  \href{http://xxx.lanl.gov/abs/astro-ph/0407529}{{\normalfont
  [arXiv:astro-ph/astro-ph/0407529]}}.
\newblock
  doi:{\changeurlcolor{black}\href{https://doi.org/10.1103/PhysRevD.70.124015}{\detokenize{10.1103/PhysRevD.70.124015}}}.

\bibitem[{Tsui} and {Leung}(2005)]{Tsui:2004qd}
{Tsui}, L.K.; {Leung}, P.T.
\newblock {Universality in quasi-normal modes of neutron stars}.
\newblock {\em Mon. Not. Roy. Astron. Soc.} {\bf 2005}, {\em 357},~1029--1037,
  \href{http://xxx.lanl.gov/abs/gr-qc/0412024}{{\normalfont
  [arXiv:gr-qc/gr-qc/0412024]}}.
\newblock
  doi:{\changeurlcolor{black}\href{https://doi.org/10.1111/j.1365-2966.2005.08710.x}{\detokenize{10.1111/j.1365-2966.2005.08710.x}}}.

\bibitem[{Lau} \em{et~al.}(2010){Lau}, {Leung}, and {Lin}]{Lau:2009bu}
{Lau}, H.K.; {Leung}, P.T.; {Lin}, L.M.
\newblock {Inferring Physical Parameters of Compact Stars from their f-mode
  Gravitational Wave Signals}.
\newblock {\em Astrophys. J.} {\bf 2010}, {\em 714},~1234--1238,
  \href{http://xxx.lanl.gov/abs/0911.0131}{{\normalfont
  [arXiv:gr-qc/0911.0131]}}.
\newblock
  doi:{\changeurlcolor{black}\href{https://doi.org/10.1088/0004-637X/714/2/1234}{\detokenize{10.1088/0004-637X/714/2/1234}}}.

\bibitem[{Bejger} and {Haensel}(2002)]{Bejger:2002ty}
{Bejger}, M.; {Haensel}, P.
\newblock {Moments of inertia for neutron and strange stars: Limits derived for
  the Crab pulsar}.
\newblock {\em Astron. Astrophys.} {\bf 2002}, {\em 396},~917--921,
  \href{http://xxx.lanl.gov/abs/astro-ph/0209151}{{\normalfont
  [arXiv:astro-ph/astro-ph/0209151]}}.
\newblock
  doi:{\changeurlcolor{black}\href{https://doi.org/10.1051/0004-6361:20021241}{\detokenize{10.1051/0004-6361:20021241}}}.

\bibitem[{Lattimer} and {Schutz}(2005)]{Lattimer:2004nj}
{Lattimer}, J.M.; {Schutz}, B.F.
\newblock {Constraining the Equation of State with Moment of Inertia
  Measurements}.
\newblock {\em Astophys. J.} {\bf 2005}, {\em 629},~979--984,
  \href{http://xxx.lanl.gov/abs/astro-ph/0411470}{{\normalfont
  [arXiv:astro-ph/astro-ph/0411470]}}.
\newblock
  doi:{\changeurlcolor{black}\href{https://doi.org/10.1086/431543}{\detokenize{10.1086/431543}}}.

\bibitem[{Urbanec} \em{et~al.}(2013){Urbanec}, {Miller}, and
  {Stuchl{\'\i}k}]{Urbanec:2013fs}
{Urbanec}, M.; {Miller}, J.C.; {Stuchl{\'\i}k}, Z.
\newblock {Quadrupole moments of rotating neutron stars and strange stars}.
\newblock {\em Mon. Not. Roy. Astron. Soc.} {\bf 2013}, {\em 433},~1903--1909,
  \href{http://xxx.lanl.gov/abs/1301.5925}{{\normalfont
  [arXiv:astro-ph.SR/1301.5925]}}.
\newblock
  doi:{\changeurlcolor{black}\href{https://doi.org/10.1093/mnras/stt858}{\detokenize{10.1093/mnras/stt858}}}.

\bibitem[{Yagi} and {Yunes}(2013{\natexlab{a}})]{Yagi:2013bca}
{Yagi}, K.; {Yunes}, N.
\newblock {I-Love-Q: Unexpected Universal Relations for Neutron Stars and Quark
  Stars}.
\newblock {\em Science} {\bf 2013}, {\em 341},~365--368,
  \href{http://xxx.lanl.gov/abs/1302.4499}{{\normalfont
  [arXiv:gr-qc/1302.4499]}}.
\newblock
  doi:{\changeurlcolor{black}\href{https://doi.org/10.1126/science.1236462}{\detokenize{10.1126/science.1236462}}}.

\bibitem[{Yagi} and {Yunes}(2013{\natexlab{b}})]{Yagi:2013awa}
{Yagi}, K.; {Yunes}, N.
\newblock {I-Love-Q relations in neutron stars and their applications to
  astrophysics, gravitational waves, and fundamental physics}.
\newblock {\em Phys. Rev. D} {\bf 2013}, {\em 88},~023009,
  \href{http://xxx.lanl.gov/abs/1303.1528}{{\normalfont
  [arXiv:gr-qc/1303.1528]}}.
\newblock
  doi:{\changeurlcolor{black}\href{https://doi.org/10.1103/PhysRevD.88.023009}{\detokenize{10.1103/PhysRevD.88.023009}}}.

\bibitem[{Yagi}(2014)]{Yagi:2013sva}
{Yagi}, K.
\newblock {Multipole Love relations}.
\newblock {\em Phys. Rev. D} {\bf 2014}, {\em 89},~043011,
  \href{http://xxx.lanl.gov/abs/1311.0872}{{\normalfont
  [arXiv:gr-qc/1311.0872]}}.
\newblock
  doi:{\changeurlcolor{black}\href{https://doi.org/10.1103/PhysRevD.89.043011}{\detokenize{10.1103/PhysRevD.89.043011}}}.

\bibitem[{Yagi} \em{et~al.}(2014){Yagi}, {Stein}, {Pappas}, {Yunes}, and
  {Apostolatos}]{Yagi:2014qua}
{Yagi}, K.; {Stein}, L.C.; {Pappas}, G.; {Yunes}, N.; {Apostolatos}, T.A.
\newblock {Why I-Love-Q: Explaining why universality emerges in compact
  objects}.
\newblock {\em Phys. Rev. D} {\bf 2014}, {\em 90},~063010,
  \href{http://xxx.lanl.gov/abs/1406.7587}{{\normalfont
  [arXiv:gr-qc/1406.7587]}}.
\newblock
  doi:{\changeurlcolor{black}\href{https://doi.org/10.1103/PhysRevD.90.063010}{\detokenize{10.1103/PhysRevD.90.063010}}}.

\bibitem[{Godzieba} \em{et~al.}(2021){Godzieba}, {Gamba}, {Radice}, and
  {Bernuzzi}]{Godzieba:2020bbz}
{Godzieba}, D.A.; {Gamba}, R.; {Radice}, D.; {Bernuzzi}, S.
\newblock {Updated universal relations for tidal deformabilities of neutron
  stars from phenomenological equations of state}.
\newblock {\em Phys. Rev. D} {\bf 2021}, {\em 103},~063036,
  \href{http://xxx.lanl.gov/abs/2012.12151}{{\normalfont
  [arXiv:astro-ph.HE/2012.12151]}}.
\newblock
  doi:{\changeurlcolor{black}\href{https://doi.org/10.1103/PhysRevD.103.063036}{\detokenize{10.1103/PhysRevD.103.063036}}}.

\bibitem[{Hinderer} \em{et~al.}(2010){Hinderer}, {Lackey}, {Lang}, and
  {Read}]{Hinderer:2009ca}
{Hinderer}, T.; {Lackey}, B.D.; {Lang}, R.N.; {Read}, J.S.
\newblock {Tidal deformability of neutron stars with realistic equations of
  state and their gravitational wave signatures in binary inspiral}.
\newblock {\em Phys. Rev. D} {\bf 2010}, {\em 81},~123016,
  \href{http://xxx.lanl.gov/abs/0911.3535}{{\normalfont
  [arXiv:astro-ph.HE/0911.3535]}}.
\newblock
  doi:{\changeurlcolor{black}\href{https://doi.org/10.1103/PhysRevD.81.123016}{\detokenize{10.1103/PhysRevD.81.123016}}}.

\bibitem[{Hinderer}(2008)]{Hinderer:2007mb}
{Hinderer}, T.
\newblock {Tidal Love Numbers of Neutron Stars}.
\newblock {\em Astrophys. J.} {\bf 2008}, {\em 677},~1216--1220,
  \href{http://xxx.lanl.gov/abs/0711.2420}{{\normalfont
  [arXiv:astro-ph/0711.2420]}}.
\newblock
  doi:{\changeurlcolor{black}\href{https://doi.org/10.1086/533487}{\detokenize{10.1086/533487}}}.

\bibitem[{Damour} and {Nagar}(2009)]{Damour:2009vw}
{Damour}, T.; {Nagar}, A.
\newblock {Relativistic tidal properties of neutron stars}.
\newblock {\em Phys. Rev. D} {\bf 2009}, {\em 80},~084035,
  \href{http://xxx.lanl.gov/abs/0906.0096}{{\normalfont
  [arXiv:gr-qc/0906.0096]}}.
\newblock
  doi:{\changeurlcolor{black}\href{https://doi.org/10.1103/PhysRevD.80.084035}{\detokenize{10.1103/PhysRevD.80.084035}}}.

\bibitem[{Nagar} \em{et~al.}(2019){Nagar}, {Messina}, {Rettegno}, {Bini},
  {Damour}, {Geralico}, {Akcay}, and {Bernuzzi}]{Nagar:2018plt}
{Nagar}, A.; {Messina}, F.; {Rettegno}, P.; {Bini}, D.; {Damour}, T.;
  {Geralico}, A.; {Akcay}, S.; {Bernuzzi}, S.
\newblock {Nonlinear-in-spin effects in effective-one-body waveform models of
  spin-aligned, inspiralling, neutron star binaries}.
\newblock {\em Phys. Rev. D} {\bf 2019}, {\em 99},~044007,
  \href{http://xxx.lanl.gov/abs/1812.07923}{{\normalfont
  [arXiv:gr-qc/1812.07923]}}.
\newblock
  doi:{\changeurlcolor{black}\href{https://doi.org/10.1103/PhysRevD.99.044007}{\detokenize{10.1103/PhysRevD.99.044007}}}.

\bibitem[{Zhao} and {Lattimer}(2018)]{Zhao:2018nyf}
{Zhao}, T.; {Lattimer}, J.M.
\newblock {Tidal deformabilities and neutron star mergers}.
\newblock {\em Phys. Rev. D} {\bf 2018}, {\em 98},~063020,
  \href{http://xxx.lanl.gov/abs/1808.02858}{{\normalfont
  [arXiv:astro-ph.HE/1808.02858]}}.
\newblock
  doi:{\changeurlcolor{black}\href{https://doi.org/10.1103/PhysRevD.98.063020}{\detokenize{10.1103/PhysRevD.98.063020}}}.

\bibitem[{Yagi} and {Yunes}(2017)]{Yagi:2016bkt}
{Yagi}, K.; {Yunes}, N.
\newblock {Approximate universal relations for neutron stars and quark stars}.
\newblock {\em Phys. Rep.} {\bf 2017}, {\em 681},~1--72,
  \href{http://xxx.lanl.gov/abs/1608.02582}{{\normalfont
  [arXiv:gr-qc/1608.02582]}}.
\newblock
  doi:{\changeurlcolor{black}\href{https://doi.org/10.1016/j.physrep.2017.03.002}{\detokenize{10.1016/j.physrep.2017.03.002}}}.

\bibitem[{Gamba} \em{et~al.}(2020){Gamba}, {Breschi}, {Bernuzzi}, {Agathos},
  and {Nagar}]{Gamba:2020wgg}
{Gamba}, R.; {Breschi}, M.; {Bernuzzi}, S.; {Agathos}, M.; {Nagar}, A.
\newblock {Waveform systematics in the gravitational-wave inference of tidal
  parameters and equation of state from binary neutron star signals}.
\newblock {\em arXiv e-prints} {\bf 2020}, p. arXiv:2009.08467,
  \href{http://xxx.lanl.gov/abs/2009.08467}{{\normalfont
  [arXiv:gr-qc/2009.08467]}}.

\bibitem[{Adhikari}(2014)]{Adhikari:2013kya}
{Adhikari}, R.X.
\newblock {Gravitational radiation detection with laser interferometry}.
\newblock {\em Reviews of Modern Physics} {\bf 2014}, {\em 86},~121--151,
  \href{http://xxx.lanl.gov/abs/1305.5188}{{\normalfont
  [arXiv:gr-qc/1305.5188]}}.
\newblock
  doi:{\changeurlcolor{black}\href{https://doi.org/10.1103/RevModPhys.86.121}{\detokenize{10.1103/RevModPhys.86.121}}}.

\bibitem[ET-()]{ET-GW}
Einstein Telescope.
\newblock \url{http://www.et-gw.eu/}.

\bibitem[{Hild} \em{et~al.}(2008){Hild}, {Chelkowski}, and
  {Freise}]{Hild:2008ng}
{Hild}, S.; {Chelkowski}, S.; {Freise}, A.
\newblock {Pushing towards the ET sensitivity using 'conventional' technology}.
\newblock {\em arXiv e-prints} {\bf 2008}, p. arXiv:0810.0604,
  \href{http://xxx.lanl.gov/abs/0810.0604}{{\normalfont
  [arXiv:gr-qc/0810.0604]}}.

\bibitem[{Reitze} \em{et~al.}(2019){Reitze}, {Adhikari}, {Ballmer}, {Barish},
  {Barsotti}, {Billingsley}, {Brown}, {Chen}, {Coyne}, {Eisenstein}, {Evans},
  {Fritschel}, {Hall}, {Lazzarini}, {Lovelace}, {Read}, {Sathyaprakash},
  {Shoemaker}, {Smith}, {Torrie}, {Vitale}, {Weiss}, {Wipf}, and
  {Zucker}]{Reitze:2019iox}
{Reitze}, D.; {Adhikari}, R.X.; {Ballmer}, S.; {Barish}, B.; {Barsotti}, L.;
  {Billingsley}, G.; {Brown}, D.A.; {Chen}, Y.; {Coyne}, D.; {Eisenstein}, R.;
  {Evans}, M.; {Fritschel}, P.; {Hall}, E.D.; {Lazzarini}, A.; {Lovelace}, G.;
  {Read}, J.; {Sathyaprakash}, B.S.; {Shoemaker}, D.; {Smith}, J.; {Torrie},
  C.; {Vitale}, S.; {Weiss}, R.; {Wipf}, C.; {Zucker}, M.
\newblock {Cosmic Explorer: The U.S. Contribution to Gravitational-Wave
  Astronomy beyond LIGO}.
\newblock  Bulletin of the American Astronomical Society,  2019, Vol.~51,
  p.~35,  \href{http://xxx.lanl.gov/abs/1907.04833}{{\normalfont
  [arXiv:astro-ph.IM/1907.04833]}}.

\bibitem[{Flanagan} and {Hinderer}(2008)]{Flanagan:2007ix}
{Flanagan}, {\'E}.{\'E}.; {Hinderer}, T.
\newblock {Constraining neutron-star tidal Love numbers with gravitational-wave
  detectors}.
\newblock {\em Phys. Rev. D} {\bf 2008}, {\em 77},~021502,
  \href{http://xxx.lanl.gov/abs/0709.1915}{{\normalfont
  [arXiv:astro-ph/0709.1915]}}.
\newblock
  doi:{\changeurlcolor{black}\href{https://doi.org/10.1103/PhysRevD.77.021502}{\detokenize{10.1103/PhysRevD.77.021502}}}.

\bibitem[{Damour} \em{et~al.}(2012){Damour}, {Nagar}, and
  {Villain}]{Damour:2012yf}
{Damour}, T.; {Nagar}, A.; {Villain}, L.
\newblock {Measurability of the tidal polarizability of neutron stars in
  late-inspiral gravitational-wave signals}.
\newblock {\em Phys. Rev. D} {\bf 2012}, {\em 85},~123007,
  \href{http://xxx.lanl.gov/abs/1203.4352}{{\normalfont
  [arXiv:gr-qc/1203.4352]}}.
\newblock
  doi:{\changeurlcolor{black}\href{https://doi.org/10.1103/PhysRevD.85.123007}{\detokenize{10.1103/PhysRevD.85.123007}}}.

\bibitem[{De} \em{et~al.}(2018){De}, {Finstad}, {Lattimer}, {Brown}, {Berger},
  and {Biwer}]{De:2018uhw}
{De}, S.; {Finstad}, D.; {Lattimer}, J.M.; {Brown}, D.A.; {Berger}, E.;
  {Biwer}, C.M.
\newblock {Tidal Deformabilities and Radii of Neutron Stars from the
  Observation of GW170817}.
\newblock {\em Phys. Rev. Lett.} {\bf 2018}, {\em 121},~091102,
  \href{http://xxx.lanl.gov/abs/1804.08583}{{\normalfont
  [arXiv:astro-ph.HE/1804.08583]}}.
\newblock
  doi:{\changeurlcolor{black}\href{https://doi.org/10.1103/PhysRevLett.121.091102}{\detokenize{10.1103/PhysRevLett.121.091102}}}.

\bibitem[{Postnikov} \em{et~al.}(2010){Postnikov}, {Prakash}, and
  {Lattimer}]{Postnikov:2010yn}
{Postnikov}, S.; {Prakash}, M.; {Lattimer}, J.M.
\newblock {Tidal Love numbers of neutron and self-bound quark stars}.
\newblock {\em Phys. Rev. D} {\bf 2010}, {\em 82},~024016,
  \href{http://xxx.lanl.gov/abs/1004.5098}{{\normalfont
  [arXiv:astro-ph.SR/1004.5098]}}.
\newblock
  doi:{\changeurlcolor{black}\href{https://doi.org/10.1103/PhysRevD.82.024016}{\detokenize{10.1103/PhysRevD.82.024016}}}.

\bibitem[{Yagi} and {Yunes}(2016)]{Yagi:2015pkc}
{Yagi}, K.; {Yunes}, N.
\newblock {Binary Love relations}.
\newblock {\em Classical and Quantum Gravity} {\bf 2016}, {\em 33},~13LT01,
  \href{http://xxx.lanl.gov/abs/1512.02639}{{\normalfont
  [arXiv:gr-qc/1512.02639]}}.
\newblock
  doi:{\changeurlcolor{black}\href{https://doi.org/10.1088/0264-9381/33/13/13LT01}{\detokenize{10.1088/0264-9381/33/13/13LT01}}}.

\bibitem[{Maselli} \em{et~al.}(2013){Maselli}, {Cardoso}, {Ferrari},
  {Gualtieri}, and {Pani}]{Maselli:2013mva}
{Maselli}, A.; {Cardoso}, V.; {Ferrari}, V.; {Gualtieri}, L.; {Pani}, P.
\newblock {Equation-of-state-independent relations in neutron stars}.
\newblock {\em Phys. Rev. D} {\bf 2013}, {\em 88},~023007,
  \href{http://xxx.lanl.gov/abs/1304.2052}{{\normalfont
  [arXiv:gr-qc/1304.2052]}}.
\newblock
  doi:{\changeurlcolor{black}\href{https://doi.org/10.1103/PhysRevD.88.023007}{\detokenize{10.1103/PhysRevD.88.023007}}}.

\bibitem[Abbott \em{et~al.}(2017)Abbott et~al.]{TheLIGOScientific:2017qsa}
Abbott, B.; others.
\newblock {GW170817: Observation of Gravitational Waves from a Binary Neutron
  Star Inspiral}.
\newblock {\em Phys. Rev. Lett.} {\bf 2017}, {\em 119},~161101,
  \href{http://xxx.lanl.gov/abs/1710.05832}{{\normalfont
  [arXiv:gr-qc/1710.05832]}}.
\newblock
  doi:{\changeurlcolor{black}\href{https://doi.org/10.1103/PhysRevLett.119.161101}{\detokenize{10.1103/PhysRevLett.119.161101}}}.

\bibitem[Abbott \em{et~al.}(2018)Abbott et~al.]{Abbott:2018exr}
Abbott, B.; others.
\newblock {GW170817: Measurements of Neutron Star Radii and Equation of State}.
\newblock {\em Phys. Rev. Lett.} {\bf 2018}, {\em 121},~161101,
  \href{http://xxx.lanl.gov/abs/1805.11581}{{\normalfont
  [arXiv:gr-qc/1805.11581]}}.
\newblock
  doi:{\changeurlcolor{black}\href{https://doi.org/10.1103/PhysRevLett.121.161101}{\detokenize{10.1103/PhysRevLett.121.161101}}}.

\bibitem[{Essick} \em{et~al.}(2020){Essick}, {Landry}, and
  {Holz}]{Essick:2019ldf}
{Essick}, R.; {Landry}, P.; {Holz}, D.E.
\newblock {Nonparametric inference of neutron star composition, equation of
  state, and maximum mass with GW170817}.
\newblock {\em Phys. Rev. D} {\bf 2020}, {\em 101},~063007,
  \href{http://xxx.lanl.gov/abs/1910.09740}{{\normalfont
  [arXiv:astro-ph.HE/1910.09740]}}.
\newblock
  doi:{\changeurlcolor{black}\href{https://doi.org/10.1103/PhysRevD.101.063007}{\detokenize{10.1103/PhysRevD.101.063007}}}.

\bibitem[{Read} \em{et~al.}(2009){Read}, {Lackey}, {Owen}, and
  {Friedman}]{Read:2008iy}
{Read}, J.S.; {Lackey}, B.D.; {Owen}, B.J.; {Friedman}, J.L.
\newblock {Constraints on a phenomenologically parametrized neutron-star
  equation of state}.
\newblock {\em Phys. Rev. D} {\bf 2009}, {\em 79},~124032,
  \href{http://xxx.lanl.gov/abs/0812.2163}{{\normalfont
  [arXiv:astro-ph/0812.2163]}}.
\newblock
  doi:{\changeurlcolor{black}\href{https://doi.org/10.1103/PhysRevD.79.124032}{\detokenize{10.1103/PhysRevD.79.124032}}}.

\bibitem[Bernuzzi and Nagar(2008)]{Bernuzzi:2008fu}
Bernuzzi, S.; Nagar, A.
\newblock {Gravitational waves from pulsations of neutron stars described by
  realistic Equations of State}.
\newblock {\em Phys. Rev.} {\bf 2008}, {\em D78},~024024,
  \href{http://xxx.lanl.gov/abs/0803.3804}{{\normalfont
  [arXiv:gr-qc/0803.3804]}}.
\newblock
  doi:{\changeurlcolor{black}\href{https://doi.org/10.1103/PhysRevD.78.024024}{\detokenize{10.1103/PhysRevD.78.024024}}}.

\bibitem[Demorest \em{et~al.}(2010)Demorest, Pennucci, Ransom, Roberts, and
  Hessels]{Demorest:2010bx}
Demorest, P.; Pennucci, T.; Ransom, S.; Roberts, M.; Hessels, J.
\newblock {Shapiro Delay Measurement of A Two Solar Mass Neutron Star}.
\newblock {\em Nature} {\bf 2010}, {\em 467},~1081--1083,
  \href{http://xxx.lanl.gov/abs/1010.5788}{{\normalfont
  [arXiv:astro-ph.HE/1010.5788]}}.
\newblock
  doi:{\changeurlcolor{black}\href{https://doi.org/10.1038/nature09466}{\detokenize{10.1038/nature09466}}}.

\bibitem[Fonseca \em{et~al.}(2016)Fonseca et~al.]{Fonseca:2016tux}
Fonseca, E.; others.
\newblock {The NANOGrav Nine-year Data Set: Mass and Geometric Measurements of
  Binary Millisecond Pulsars}.
\newblock {\em Astrophys. J.} {\bf 2016}, {\em 832},~167,
  \href{http://xxx.lanl.gov/abs/1603.00545}{{\normalfont
  [arXiv:astro-ph.HE/1603.00545]}}.
\newblock
  doi:{\changeurlcolor{black}\href{https://doi.org/10.3847/0004-637X/832/2/167}{\detokenize{10.3847/0004-637X/832/2/167}}}.

\bibitem[Arzoumanian \em{et~al.}(2018)Arzoumanian et~al.]{Arzoumanian:2017puf}
Arzoumanian, Z.; others.
\newblock {The NANOGrav 11-year Data Set: High-precision timing of 45
  Millisecond Pulsars}.
\newblock {\em Astrophys. J. Suppl.} {\bf 2018}, {\em 235},~37,
  \href{http://xxx.lanl.gov/abs/1801.01837}{{\normalfont
  [arXiv:astro-ph.HE/1801.01837]}}.
\newblock
  doi:{\changeurlcolor{black}\href{https://doi.org/10.3847/1538-4365/aab5b0}{\detokenize{10.3847/1538-4365/aab5b0}}}.

\bibitem[Antoniadis \em{et~al.}(2013)Antoniadis et~al.]{Antoniadis:2013pzd}
Antoniadis, J.; others.
\newblock {A Massive Pulsar in a Compact Relativistic Binary}.
\newblock {\em Science} {\bf 2013}, {\em 340},~6131,
  \href{http://xxx.lanl.gov/abs/1304.6875}{{\normalfont
  [arXiv:astro-ph.HE/1304.6875]}}.
\newblock
  doi:{\changeurlcolor{black}\href{https://doi.org/10.1126/science.1233232}{\detokenize{10.1126/science.1233232}}}.

\bibitem[Cromartie \em{et~al.}(2019)Cromartie et~al.]{Cromartie:2019kug}
Cromartie, H.T.; others.
\newblock {Relativistic Shapiro delay measurements of an extremely massive
  millisecond pulsar}.
\newblock {\em Nat. Astron.} {\bf 2019}, {\em 4},~72--76,
  \href{http://xxx.lanl.gov/abs/1904.06759}{{\normalfont
  [arXiv:astro-ph.HE/1904.06759]}}.
\newblock
  doi:{\changeurlcolor{black}\href{https://doi.org/10.1038/s41550-019-0880-2}{\detokenize{10.1038/s41550-019-0880-2}}}.

\bibitem[Linares \em{et~al.}(2018)Linares, Shahbaz, and
  Casares]{Linares:2018ppq}
Linares, M.; Shahbaz, T.; Casares, J.
\newblock {Peering into the dark side: Magnesium lines establish a massive
  neutron star in PSR J2215+5135}.
\newblock {\em Astrophys. J.} {\bf 2018}, {\em 859},~54,
  \href{http://xxx.lanl.gov/abs/1805.08799}{{\normalfont
  [arXiv:astro-ph.HE/1805.08799]}}.
\newblock
  doi:{\changeurlcolor{black}\href{https://doi.org/10.3847/1538-4357/aabde6}{\detokenize{10.3847/1538-4357/aabde6}}}.

\bibitem[Margalit and Metzger(2017)]{Margalit:2017dij}
Margalit, B.; Metzger, B.D.
\newblock {Constraining the Maximum Mass of Neutron Stars From Multi-Messenger
  Observations of GW170817}.
\newblock {\em Astrophys. J. Lett.} {\bf 2017}, {\em 850},~L19,
  \href{http://xxx.lanl.gov/abs/1710.05938}{{\normalfont
  [arXiv:astro-ph.HE/1710.05938]}}.
\newblock
  doi:{\changeurlcolor{black}\href{https://doi.org/10.3847/2041-8213/aa991c}{\detokenize{10.3847/2041-8213/aa991c}}}.

\bibitem[Shibata \em{et~al.}(2017)Shibata, Fujibayashi, Hotokezaka, Kiuchi,
  Kyutoku, Sekiguchi, and Tanaka]{Shibata:2017xdx}
Shibata, M.; Fujibayashi, S.; Hotokezaka, K.; Kiuchi, K.; Kyutoku, K.;
  Sekiguchi, Y.; Tanaka, M.
\newblock {Modeling GW170817 based on numerical relativity and its
  implications}.
\newblock {\em Phys. Rev.} {\bf 2017}, {\em D96},~123012,
  \href{http://xxx.lanl.gov/abs/1710.07579}{{\normalfont
  [arXiv:astro-ph.HE/1710.07579]}}.
\newblock
  doi:{\changeurlcolor{black}\href{https://doi.org/10.1103/PhysRevD.96.123012}{\detokenize{10.1103/PhysRevD.96.123012}}}.

\bibitem[Rezzolla \em{et~al.}(2018)Rezzolla, Most, and Weih]{Rezzolla:2017aly}
Rezzolla, L.; Most, E.R.; Weih, L.R.
\newblock {Using gravitational-wave observations and quasi-universal relations
  to constrain the maximum mass of neutron stars}.
\newblock {\em Astrophys. J. Lett.} {\bf 2018}, {\em 852},~L25,
  \href{http://xxx.lanl.gov/abs/1711.00314}{{\normalfont
  [arXiv:astro-ph.HE/1711.00314]}}.
\newblock
  doi:{\changeurlcolor{black}\href{https://doi.org/10.3847/2041-8213/aaa401}{\detokenize{10.3847/2041-8213/aaa401}}}.

\bibitem[Ruiz \em{et~al.}(2018)Ruiz, Shapiro, and Tsokaros]{Ruiz:2017due}
Ruiz, M.; Shapiro, S.L.; Tsokaros, A.
\newblock {GW170817, General Relativistic Magnetohydrodynamic Simulations, and
  the Neutron Star Maximum Mass}.
\newblock {\em Phys. Rev.} {\bf 2018}, {\em D97},~021501,
  \href{http://xxx.lanl.gov/abs/1711.00473}{{\normalfont
  [arXiv:astro-ph.HE/1711.00473]}}.
\newblock
  doi:{\changeurlcolor{black}\href{https://doi.org/10.1103/PhysRevD.97.021501}{\detokenize{10.1103/PhysRevD.97.021501}}}.

\bibitem[Shibata \em{et~al.}(2019)Shibata, Zhou, Kiuchi, and
  Fujibayashi]{Shibata:2019ctb}
Shibata, M.; Zhou, E.; Kiuchi, K.; Fujibayashi, S.
\newblock {Constraint on the maximum mass of neutron stars using GW170817
  event}.
\newblock {\em Phys. Rev.} {\bf 2019}, {\em D100},~023015,
  \href{http://xxx.lanl.gov/abs/1905.03656}{{\normalfont
  [arXiv:astro-ph.HE/1905.03656]}}.
\newblock
  doi:{\changeurlcolor{black}\href{https://doi.org/10.1103/PhysRevD.100.023015}{\detokenize{10.1103/PhysRevD.100.023015}}}.

\bibitem[{Godzieba} \em{et~al.}(2021){Godzieba}, {Radice}, and
  {Bernuzzi}]{Godzieba:2020tjn}
{Godzieba}, D.A.; {Radice}, D.; {Bernuzzi}, S.
\newblock {On the Maximum Mass of Neutron Stars and GW190814}.
\newblock {\em Astrophys. J.} {\bf 2021}, {\em 908},~122,
  \href{http://xxx.lanl.gov/abs/2007.10999}{{\normalfont
  [arXiv:astro-ph.HE/2007.10999]}}.
\newblock
  doi:{\changeurlcolor{black}\href{https://doi.org/10.3847/1538-4357/abd4dd}{\detokenize{10.3847/1538-4357/abd4dd}}}.

\bibitem[{Martinez} \em{et~al.}(2015){Martinez}, {Stovall}, {Freire}, {Deneva},
  {Jenet}, {McLaughlin}, {Bagchi}, {Bates}, and {Ridolfi}]{Martinez:2015mya}
{Martinez}, J.G.; {Stovall}, K.; {Freire}, P.C.C.; {Deneva}, J.S.; {Jenet},
  F.A.; {McLaughlin}, M.A.; {Bagchi}, M.; {Bates}, S.D.; {Ridolfi}, A.
\newblock {Pulsar J0453+1559: A Double Neutron Star System with a Large Mass
  Asymmetry}.
\newblock {\em Astrophys. J.} {\bf 2015}, {\em 812},~143,
  \href{http://xxx.lanl.gov/abs/1509.08805}{{\normalfont
  [arXiv:astro-ph.HE/1509.08805]}}.
\newblock
  doi:{\changeurlcolor{black}\href{https://doi.org/10.1088/0004-637X/812/2/143}{\detokenize{10.1088/0004-637X/812/2/143}}}.

\bibitem[{Suwa} \em{et~al.}(2018){Suwa}, {Yoshida}, {Shibata}, {Umeda}, and
  {Takahashi}]{Suwa:2018uni}
{Suwa}, Y.; {Yoshida}, T.; {Shibata}, M.; {Umeda}, H.; {Takahashi}, K.
\newblock {On the minimum mass of neutron stars}.
\newblock {\em Mon. Not. Roy. Astron. Soc.} {\bf 2018}, {\em 481},~3305--3312,
  \href{http://xxx.lanl.gov/abs/1808.02328}{{\normalfont
  [arXiv:astro-ph.HE/1808.02328]}}.
\newblock
  doi:{\changeurlcolor{black}\href{https://doi.org/10.1093/mnras/sty2460}{\detokenize{10.1093/mnras/sty2460}}}.

\bibitem[{Abbott} \em{et~al.}(2020){Abbott} et~al.]{KAGRA:2013rdx}
{Abbott}, B.P.; others.
\newblock {Prospects for observing and localizing gravitational-wave transients
  with Advanced LIGO, Advanced Virgo and KAGRA}.
\newblock {\em Living Reviews in Relativity} {\bf 2020}, {\em 23},~3.
\newblock
  doi:{\changeurlcolor{black}\href{https://doi.org/10.1007/s41114-020-00026-9}{\detokenize{10.1007/s41114-020-00026-9}}}.

\bibitem[{Danielewicz} \em{et~al.}(2002){Danielewicz}, {Lacey}, and
  {Lynch}]{Danielewicz:2002pu}
{Danielewicz}, P.; {Lacey}, R.; {Lynch}, W.G.
\newblock {Determination of the Equation of State of Dense Matter}.
\newblock {\em Science} {\bf 2002}, {\em 298},~1592--1596,
  \href{http://xxx.lanl.gov/abs/nucl-th/0208016}{{\normalfont
  [arXiv:nucl-th/nucl-th/0208016]}}.
\newblock
  doi:{\changeurlcolor{black}\href{https://doi.org/10.1126/science.1078070}{\detokenize{10.1126/science.1078070}}}.

\bibitem[{Reed} \em{et~al.}(2021){Reed}, {Fattoyev}, {Horowitz}, and
  {Piekarewicz}]{Reed:2021nqk}
{Reed}, B.T.; {Fattoyev}, F.J.; {Horowitz}, C.J.; {Piekarewicz}, J.
\newblock {Implications of PREX-2 on the Equation of State of Neutron-Rich
  Matter}.
\newblock {\em Phys. Rev. Lett.} {\bf 2021}, {\em 126},~172503,
  \href{http://xxx.lanl.gov/abs/2101.03193}{{\normalfont
  [arXiv:nucl-th/2101.03193]}}.
\newblock
  doi:{\changeurlcolor{black}\href{https://doi.org/10.1103/PhysRevLett.126.172503}{\detokenize{10.1103/PhysRevLett.126.172503}}}.

\bibitem[{Steiner} \em{et~al.}(2010){Steiner}, {Lattimer}, and
  {Brown}]{Steiner:2010fz}
{Steiner}, A.W.; {Lattimer}, J.M.; {Brown}, E.F.
\newblock {The Equation of State from Observed Masses and Radii of Neutron
  Stars}.
\newblock {\em Astrophys. J.} {\bf 2010}, {\em 722},~33--54,
  \href{http://xxx.lanl.gov/abs/1005.0811}{{\normalfont
  [arXiv:astro-ph.HE/1005.0811]}}.
\newblock
  doi:{\changeurlcolor{black}\href{https://doi.org/10.1088/0004-637X/722/1/33}{\detokenize{10.1088/0004-637X/722/1/33}}}.

\bibitem[{Lattimer}(2012)]{Lattimer:2012nd}
{Lattimer}, J.M.
\newblock {The Nuclear Equation of State and Neutron Star Masses}.
\newblock {\em Annual Review of Nuclear and Particle Science} {\bf 2012}, {\em
  62},~485--515,  \href{http://xxx.lanl.gov/abs/1305.3510}{{\normalfont
  [arXiv:nucl-th/1305.3510]}}.
\newblock
  doi:{\changeurlcolor{black}\href{https://doi.org/10.1146/annurev-nucl-102711-095018}{\detokenize{10.1146/annurev-nucl-102711-095018}}}.

\bibitem[{{\"O}zel} and {Freire}(2016)]{Ozel:2016oaf}
{{\"O}zel}, F.; {Freire}, P.
\newblock {Masses, Radii, and the Equation of State of Neutron Stars}.
\newblock {\em Ann. Rev. Astro. Astrophys.} {\bf 2016}, {\em 54},~401--440,
  \href{http://xxx.lanl.gov/abs/1603.02698}{{\normalfont
  [arXiv:astro-ph.HE/1603.02698]}}.
\newblock
  doi:{\changeurlcolor{black}\href{https://doi.org/10.1146/annurev-astro-081915-023322}{\detokenize{10.1146/annurev-astro-081915-023322}}}.

\bibitem[{Annala} \em{et~al.}(2020){Annala}, {Gorda}, {Kurkela},
  {N{\"a}ttil{\"a}}, and {Vuorinen}]{Annala:2019puf}
{Annala}, E.; {Gorda}, T.; {Kurkela}, A.; {N{\"a}ttil{\"a}}, J.; {Vuorinen}, A.
\newblock {Evidence for quark-matter cores in massive neutron stars}.
\newblock {\em Nature Physics} {\bf 2020}, {\em 16},~907--910,
  \href{http://xxx.lanl.gov/abs/1903.09121}{{\normalfont
  [arXiv:astro-ph.HE/1903.09121]}}.
\newblock
  doi:{\changeurlcolor{black}\href{https://doi.org/10.1038/s41567-020-0914-9}{\detokenize{10.1038/s41567-020-0914-9}}}.

\bibitem[Godzieba \em{et~al.}(2020)Godzieba, Radice, and
  Bernuzzi]{godzieba_daniel_a_2020_3954899}
Godzieba, D.A.; Radice, D.; Bernuzzi, S.
\newblock Phenomenological EOS Data Set,  2020.
\newblock
  doi:{\changeurlcolor{black}\href{https://doi.org/10.5281/zenodo.3954899}{\detokenize{10.5281/zenodo.3954899}}}.

\end{thebibliography}

%


\end{document}